\documentclass[lettersize,journal]{IEEEtran}
\IEEEoverridecommandlockouts
\usepackage{cite}
\usepackage{amsmath,amssymb,amsfonts}
\usepackage{multirow}
\usepackage{algorithm}
\usepackage{algpseudocode}
\usepackage{graphicx}
\usepackage{textcomp}
\usepackage{xcolor}
\usepackage[numbers]{natbib}
\usepackage{hyperref}
\usepackage{graphicx}
\usepackage{subcaption}
\usepackage{tabularray}
\hypersetup{
 colorlinks=true,
 linkcolor=blue,
 filecolor=blue,      
 urlcolor=blue,
 citecolor=blue,
}

\def\BibTeX{{\rm B\kern-.05em{\sc i\kern-.025em b}\kern-.08em
    T\kern-.1667em\lower.7ex\hbox{E}\kern-.125emX}}

\begin{document}

\title{EAT: QoS-Aware Edge-Collaborative AIGC Task Scheduling via Attention-Guided Diffusion Reinforcement Learning}

\author{Zhifei Xu, Zhiqing Tang,~\IEEEmembership{Member,~IEEE}, Jiong Lou,~\IEEEmembership{Member,~IEEE}, Zhi Yao, Xuan Xie,\\Tian Wang,~\IEEEmembership{Senior Member,~IEEE}, Yinglong Wang,~\IEEEmembership{Senior Member,~IEEE}, and Weijia Jia,~\IEEEmembership{Fellow,~IEEE}
\thanks{Zhifei Xu and Xuan Xie are with Faculty of Arts and Sciences, Beijing Normal University, Zhuhai 519087, China, and also with Institute of Artificial Intelligence and Future Networks, Beijing Normal University, Zhuhai 519087, China (e-mail: \{xuzhifei, xuanxie\}@mail.bnu.edu.cn).}
\thanks{Zhiqing Tang is with Institute of Artificial Intelligence and Future Networks, Beijing Normal University, Zhuhai 519087, China, and also with Key Laboratory of Computing Power Network and Information Security, Ministry of Education, Shandong Computer Science Center, Qilu University of Technology (Shandong Academy of Sciences), Jinan 250014, China (e-mail: zhiqingtang@bnu.edu.cn).}
\thanks{Jiong Lou is with Department of Computer Science and Engineering, Shanghai Jiao Tong University, Shanghai 200240, China (e-mail: lj1994@sjtu.edu.cn).}
\thanks{Zhi Yao is with School of Artificial Intelligence, Beijing Normal University, Beijing 100875, China, and also with Institute of Artificial Intelligence and Future Networks, Beijing Normal University, Zhuhai 519087, China (e-mail: yaozhi@mail.bnu.edu.cn).}
\thanks{Tian Wang is with Institute of Artificial Intelligence and Future Networks, Beijing Normal University, Zhuhai 519087, China (e-mail: tianwang@bnu.edu.cn).}
\thanks{Yinglong Wang is with Key Laboratory of Computing Power Network and Information Security, Ministry of Education, Shandong Computer Science Center, Qilu University of Technology (Shandong Academy of Sciences), Jinan 250014, China (e-mail: wangyinglong@qlu.edu.cn).}
\thanks{Weijia Jia is with Institute of Artificial Intelligence and Future Networks, Beijing Normal University, Zhuhai 519087, China and also with Guangdong Key Lab of AI and Multi-Modal Data Processing, Beijing Normal-Hong Kong Baptist University, Zhuhai 519087, China (e-mail: jiawj@bnu.edu.cn).}
}

\maketitle

\begin{abstract}

The growth of Artificial Intelligence (AI) and large language models has enabled the use of Generative AI (GenAI) in cloud data centers for diverse AI-Generated Content (AIGC) tasks. Models like Stable Diffusion introduce unavoidable delays and substantial resource overhead, which are unsuitable for users at the network edge with high QoS demands. Deploying AIGC services on edge servers reduces transmission times but often leads to underutilized resources and fails to optimally balance inference latency and quality. To address these issues, this paper introduces a QoS-aware \underline{E}dge-collaborative \underline{A}IGC \underline{T}ask scheduling (EAT) algorithm. Specifically: 1) We segment AIGC tasks and schedule patches to various edge servers, formulating it as a gang scheduling problem that balances inference latency and quality while considering server heterogeneity, such as differing model distributions and cold start issues. 2) We propose a reinforcement learning-based EAT algorithm that uses an attention layer to extract load and task queue information from edge servers and employs a diffusion-based policy network for scheduling, efficiently enabling model reuse. 3) We develop an AIGC task scheduling system that uses our EAT algorithm to divide tasks and distribute them across multiple edge servers for processing. Experimental results based on our system and large-scale simulations show that our EAT algorithm can reduce inference latency by up to 56\% compared to baselines. We release our open-source code at https://github.com/zzf1955/EAT.

\end{abstract}


\section{Introduction} \label{Introduction}

\begin{figure*}[!t]
    \centering
    \includegraphics[width=0.97\linewidth, page=1]{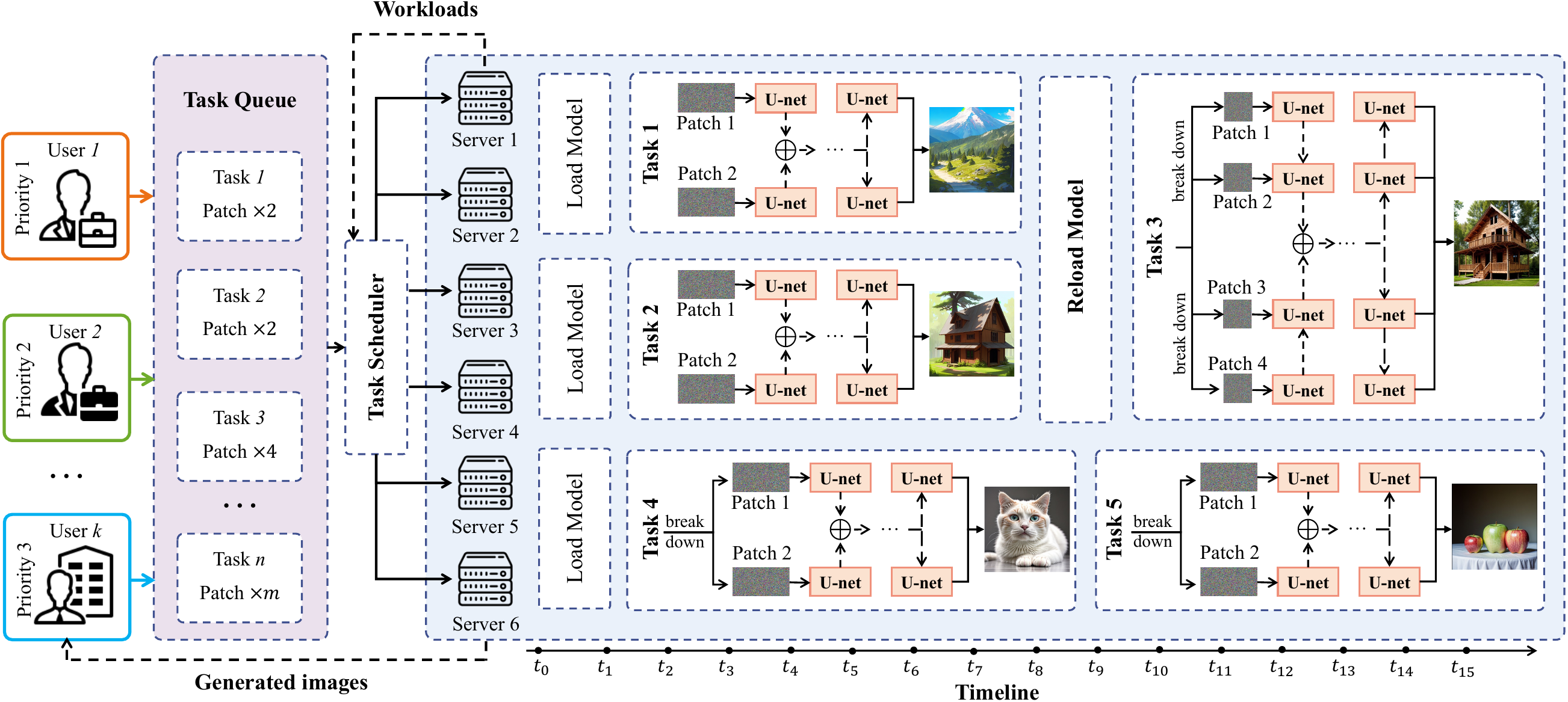} 
    \caption{Example of QoS-aware edge-collaborative AIGC task scheduling. Tasks with different priorities will be divided into different numbers of patches and scheduled accordingly. For example, Task $1$ from User $1$ is deployed on Server $1$ and Server $2$, while Task $2$ is deployed on Server $3$ and Server $4$. Subsequently, Servers $1-4$ reload the model to execute Task $3$, which featured a different AIGC service configuration. Server load, inference steps, and model conditions impact AIGC task execution and initialization times. Effective scheduling must balance quality and latency while accounting for these factors.}
    \label{Gang-Scheduling}
\end{figure*}


The rapid development of Generative AI (GenAI) and the proliferation of large-scale AI models, such as Diffusion Models (DMs) \cite{b3.2} and Large Language Models (LLMs) \cite{b3.3}, have greatly increased the computational demands of AI-Generated Content (AIGC) tasks like image and text generation. Local devices often struggle to handle these AIGC tasks efficiently due to limited resources. Therefore, AIGC services are typically deployed on distributed systems, such as cloud computing and edge computing \cite{b1.2, wang2025empowering}. Edge servers closer to users reduce transmission times, while more distant servers or remote clouds offer greater computing power to minimize inference time. This has motivated research on deploying AIGC services across multiple edge servers and selecting the best service providers~\cite{b2.3} to reduce inference latency \cite{b2.4, b3.13, tang2024multi}, enhance resource utilization \cite{b2.13, b2.20}, and reduce energy consumption \cite{b2.3, b3.5}.


However, generating high-quality images with GenAI requires a considerable number of inference steps, leading to significant inference time. Therefore, it necessitates optimizations beyond just server selection \cite{b2.3}. Some training-free methods are proposed to reduce the latency \cite{ma2024learning, ye2024training}. Besides, DistriFusion \cite{b3.4} addresses this by dividing AIGC tasks into several patches and distributing them across multiple servers for parallel processing, effectively reducing inference latency without compromising quality. The issue remains in efficiently scheduling these patches across edge servers or the cloud to balance inference latency and quality. For example, for high-priority tasks with sufficient resources, splitting and parallel processing can greatly reduce inference latency. Under resource constraints, consolidating tasks on a single server with minimal quality trade-offs may be more efficient \cite{b3.12}. Furthermore, for various tasks, reusing existing models and minimizing reloads can help reduce cold start latency. Therefore, resource-aware scheduling of AIGC tasks through edge collaboration to balance inference latency and Quality of Service (QoS) remains a critical issue.

To fill in such gaps, this paper presents the design and implementation of an edge-collaborative AIGC task scheduling system for multiple and distributed AIGC models. As illustrated in Fig.~\ref{Gang-Scheduling}, tasks from users are categorized into patches. Each task's multiple patches require a selection of servers to ensure simultaneous processing. Given that different AIGC tasks necessitate loading distinct models, which incurs considerable initialization time \cite{wan2023efficient}, scheduling must consider factors such as the resource availability of edge servers and the models in use. In summary, to achieve QoS-aware edge-collaborative AIGC task scheduling, the following challenges must be addressed.

\textit{The first challenge is how to efficiently reuse models across heterogeneous edge servers to minimize initialization costs and time.} Large AIGC models in distributed systems experience significant initialization and switching delays \cite{wang2025empowering}, particularly when task scheduling repeatedly uses the same models on different edge servers. This frequent reloading creates substantial overhead, which can be greatly reduced by reusing already initialized models. For model reutilization, energy-based parameterization and novel samplers have been introduced to enable compositional reuse of pre-trained diffusion models, allowing them to be combined for complex generative tasks at inference without retraining \cite{du2023reduce}. In knowledge distillation, the teacher's classifier is directly reused during student inference, with the student's encoder trained to align its features to the teacher’s, eliminating the need for complex knowledge representations \cite{chen2022knowledge}. In self-training, image comprehension in large vision-language models is improved by building a preference dataset from unlabeled images and reusing instruction-tuning data augmented with self-generated descriptions \cite{deng2024enhancing}. However, the heterogeneity and distribution of edge resources lead to variations in model performance across different servers, e.g., startup time, inference time, cost. Therefore, we analyze edge server resources and propose an attention-based feature extraction method to efficiently capture resource information, facilitating model reuse across servers \cite{b4.3}.

\textit{The second challenge is how to efficiently scheduling parallel AIGC tasks across heterogeneous edge servers under a gang scheduling requirement.} To efficiently schedule distributed and parallel tasks in edge computing, some resource allocation schemes have been proposed for server selection \cite{fan2025dynamic}. An online scheduling architecture applies Reinforcement Learning (RL) to schedule dependent tasks, proactively managing and reusing executors to minimize startup delays and reduce overall task completion time \cite{zou2025reflexpilot}. A collaborative edge AI system speeds up Transformer inference by distributing tasks across diverse devices with a novel hybrid model parallelism \cite{ye2025resource}. For AIGC tasks, they can be divided into patches for simultaneous processing by multiple edge servers~\cite{b3.4}. Each patch of a single task must be dispatched and run simultaneously, placing unique demands on server resources and placement decisions~\cite{lee2022design}. However, the variety of server capabilities and multiple AIGC models complicate matters, as frequent model switching or re-initialization can cause notable latency \cite{b3.7}. To address this challenge, we model AIGC task scheduling as a gang scheduling problem, aiming to minimize task inference latency while dynamically adjusting task quality. Then, we employ a diffusion model to generate gang scheduling actions for selecting appropriate tasks and cooperative servers.

\textit{The third challenge is how to jointly determine the number of inference steps and task scheduling to optimize both inference latency and quality.} Fewer inference steps reduce inference time but compromise output quality, while more steps enhance quality at the risk of exceeding latency requirements \cite{b3.2,b3.1}. Text-to-Image generation can be optimized on edge devices by adaptively scheduling diffusion processes across different models, balancing resource allocation to ensure service quality and low latency for multiple users \cite{gao2025characterizing}. An algorithm-system co-design speeds up diffusion model inference on mobile devices by leveraging lookup tables and parallel CPU-GPU co-scheduling, adaptively choosing the optimal strategy to minimize overhead without expensive retraining \cite{wang2025efficient}. However, large AIGC models incur significant overhead when switching tasks or reusing states across different servers \cite{b3.7}. Efficient scheduling must optimize these factors while meeting QoS requirements. To tackle this issue, we propose a RL-based algorithm for dynamic scheduling decisions that considers the latency-quality trade-off. However, the action space becomes much more complex, as we need to output multiple discrete and continuous actions. Therefore, we integrate the attention-guided diffusion model into the RL policy network to enable the actor to make the complex decisions. We also use dynamic latency prediction to improve scheduling accuracy.

In this paper, we formulate the QoS-aware \underline{E}dge-collaborative \underline{A}IGC \underline{T}ask scheduling (EAT) problem, aiming to optimize the balance between inference latency and quality. We dynamically adjust the number of inference steps to respond to adapt to server load and meet QoS requirements. To enhance time efficiency, we enable model reuse for similar tasks, allowing new tasks to leverage previously loaded models. We propose the EAT algorithm to address the aforementioned challenges. Our edge system comprises multiple servers and GPUs, on which we have implemented our AIGC task scheduling and processing framework based on DistriFusion \cite{b3.4}. We have realized the entire process illustrated in Fig.~\ref{Gang-Scheduling} and evaluate the EAT algorithm against baseline algorithms in this system, along with larger-scale simulation experiments. Results indicate that EAT improves inference latency by approximately 25\% while maintaining superior task quality compared to baselines in real environments. The main contributions are summarized as follows:

\begin{enumerate}
    \item We model the AIGC task gang scheduling problem as a Markov Decision Process (MDP), accounting for the interdependencies among parallel distributed patches, model initialization overhead, and resource utilization to optimize the quality-time trade-off in AIGC tasks.
    \item We propose the EAT algorithm for AIGC task scheduling, utilizing an attention mechanism to extract key features and a diffusion model to generate high-quality scheduling actions. EAT enhances overall efficiency while maintaining content quality.
    \item We implement an edge system and conduct real-machine validation. The results show that EAT effectively addresses AIGC task scheduling through dynamic quality adjustment and reduced inference latency.
\end{enumerate}

The paper is structured as follows: Section \ref{Relatedwork} reviews related work, Section \ref{System Design and Problem Modeling} details the system design and problem formulation, Section \ref{Algorithm Design} presents the EAT algorithm, Section \ref{Experiment And Result} evaluates performance, and Section \ref{Conclusion} concludes the paper.

\section{Background and Motivation}

In GenAI models like Stable Diffusion \cite{b3.2}, inference latency and task quality depend on the number of inference steps. Stable Diffusion creates images by repeatedly denoising a noise sample, with these iterations referred to as inference steps. Within a certain range, more steps improve image quality \cite{b3.12}. However, since each step requires the denoising network to predict noise, inference latency increases proportionally with the number of steps. Large AIGC tasks can be divided into smaller patches and executed concurrently across multiple servers. Li \textit{et al.} \cite{b3.4} evaluate the acceleration of Stable Diffusion tasks split into different numbers of patches, with TABLE~\ref{tab:acceleration} showing substantial performance improvements.


\begin{table}[t]
\centering
\caption{Task Acceleration with Different Number of Patches}
\begin{tabular}{c|c|c}
\hline
\textbf{Number of Patches} & \textbf{Time (s)} & \textbf{Acceleration} \\
\hline
1& 23.7 & $\times$1 \\
2& 13.3 & $\times$1.8 \\
4& 7.6  & $\times$3.1 \\
8& 4.81 & $\times$4.9 \\
\hline
\end{tabular}
\label{tab:acceleration}
\end{table}

However, the existing research overlooks server heterogeneity, including differences in resources and model distribution, while patch scheduling strategies greatly influence AIGC task inference latency. We implement a traditional algorithm with fixed inference steps and sequential task execution \cite{b3.4}, and then compare our EAT algorithm against it on a server equipped with four NVIDIA 4090D 24G GPUs, as shown in TABLE~\ref{tab:EAT_Example} and TABLE~\ref{tab:Fixed Example}. In our experiments, tasks $1$, $2$, $3$, and $4$ are submitted sequentially with a 10-second interval between each. When task priorities are low, quality can be sacrificed for quicker completion to prevent delays in subsequent tasks. Therefore, our EAT algorithm reduces the inference steps for Tasks $1$ and $2$ in TABLE~\ref{tab:EAT_Example}. Secondly, the initialization results should be reused to avoid redundant initializations, such as reducing Init $2$ and Init $3$ in TABLE~\ref{tab:Fixed Example}. Overall, the results in TABLE~\ref{tab:algorithm_comp} show that the EAT algorithm effectively reduces task inference latency by reusing initialization and balancing time and quality, while maintaining high quality. The traditional method, although offering slightly better quality, is hindered by fixed execution steps and repeated initialization, resulting in significant delays. This highlights the need for a QoS-aware task scheduling method for AIGC tasks.

\begin{table}[t]
\centering
\caption{EAT Algorithm Example}
\label{tab:EAT_Example}
\begin{tabular}{c|c|c|c|c|c|c}
\hline
\textbf{Task} & \textbf{Patch} & \textbf{GPU} & \textbf{Step} & \textbf{Time} & \textbf{Inference (s)} & \textbf{Quality} \\ 
\hline
Init 1& - & 1 2 & - & 28.0 & - & - \\
Init 2& - & 3 4 & - & 28.0 & - & - \\
Task 1 & 2 & 1 2 & 18 & 5.2 & 33.2 & 2.4 \\
Task 2 & 2 & 3 4 & 17 & 4.9 & 22.9 & 2.4 \\
Task 4 & 2 & 1 2 & 25 & 7.3 & 10.5 & 2.7 \\
Init 3& - & 1 2 3 4 & - & 35.0 & - & - \\
Task 3 & 4 & 1 2 3 4 & 17 & 3.5 & 23.9 & 2.4 \\
\hline
\end{tabular}
\end{table}

\begin{table}[t]
\centering
\caption{Traditional Algorithm Example}
\label{tab:Fixed Example}
\begin{tabular}{c|c|c|c|c|c|c}
\hline
\textbf{Task} & \textbf{Patch} & \textbf{GPU} & \textbf{Step} & \textbf{Time} & \textbf{Inference (s)} & \textbf{Quality} \\ 
\hline
Init 1& - & 1 2 & - & 28 & - & - \\
Task 1 & 2 & 1 2 & \textbf{20} & 5.8 & 33.8 & 2.51 \\
Task 2 & 2 & 3 4 & \textbf{20} & 5.8 & 29.6 & 2.51 \\
\textbf{Init 2} & - & 1 2 3 4 & - & \textbf{35} & - & - \\
Task 3 & 4 & 1 2 3 4 & 20 & 5.8 & 60.4 & 2.51 \\
\textbf{Init 3} & - & 1 2 & - & \textbf{28} & - & - \\
Task 4 & 2 & 1 2 & 20 & 5.8 & 84.2 & 2.51 \\
\hline
\end{tabular}
\end{table}

\begin{table}[t]
\centering
\caption{Algorithm Performance Comparison}
\begin{tabular}{l|c|c}
\hline
\textbf{Metric} & \textbf{EAT} & \textbf{Traditional} \\
\hline
Quality & 2.4 & 2.51 \\
Inference Latency (s) & 22.64
 & 52.00 \\
\hline
\end{tabular}
\label{tab:algorithm_comp}
\end{table}

\section{Related Work} \label{Relatedwork}

\textit{AIGC Task.} Recent years have seen major advancements in AI, especially in AIGC, with model scales steadily growing \cite{b3.7}. Rombach \textit{et al.} \cite{b3.2} apply diffusion models to image generation. Mann \textit{et al.} \cite{b3.8} pioneer the GPT model. Jaech \textit{et al.} \cite{b3.10} introduce reasoning model o1, significantly enhancing the capabilities of large language models. Scaling AIGC models brings challenges like distributed inference and efficiency improvements \cite{b3.6}. Researchers are also addressing the energy demands of distributed inference \cite{b1.3} and the environmental impact of increasingly complex AI models \cite{b3.5}. To further optimize diffusion model inference, Li \textit{et al.} \cite{b3.4} propose a distributed diffusion method.


\textit{Diffusion in RL.} In edge computing, RL has been effectively applied to challenges such as service placement \cite{b2.6} and task offloading \cite{b2.9}. Meanwhile, diffusion models have demonstrated outstanding performance in a variety of generative tasks, like image generation \cite{b3.2} and molecular structure design \cite{b3.12}. Increasing research focuses on integrating diffusion models with RL. Wang \textit{et al.} \cite{b2.1} introduce diffusion Q-learning that combines behavioral cloning and Q-learning using a diffusion model. Fontanesi \textit{et al.} \cite{b2.11} propose the RL diffusion decision model, showcasing improved value-based decision-making dynamics. Du \textit{et al.} \cite{b2.3} develop the Deep Diffusion Soft Actor-Critic (D2SAC) algorithm to enhance AIGC services in edge computing networks.

\section{System Model and Problem Formulation} \label{System Design and Problem Modeling}

\iftrue
\begin{table}[ht]
\centering
\caption{Notations}
\label{tab:symbols}
\begin{tabular}{c|l}
\hline
\textbf{Symbol} & \textbf{Description} \\
\hline
$\mathbf{K}$ & Set of tasks to be submitted by the user. \\
$k$ & $k^{th}$ task. \\
$t^a_k$ & Arrival time of task $k$. \\
$t^s_k$ & Start time of task $k$. \\
$t^e_k$ & Predicted execution time of task $k$. \\
$t^r_e$ & Estimated remaining time of server $e$. \\
$t^g_k$ & Generation interval between tasks $k$ and $k+1$. \\
$c_k$ & Number of cooperating servers required for task $k$. \\
$s_k$ & Time steps of task $k$ executed in Stable Diffusion. \\
$S_{min}$ & Minimum threshold for inference steps. \\
$S_{max}$ & Maximum threshold for inference steps. \\
$g_k$ & Prompt of task $k$. \\
$i_k$ & Result image generated by task $k$. \\
$\mathbf{Q}$ & List of tasks in the task queue. \\
$l$ & The number of tasks can be selected. \\
$\mathbf{E}$ & Set of servers in the cluster system. \\
$\mathbf{E}_k$ & Set of servers seleted to run task $k$. \\
$e$ & $e^{th}$ server. \\
$a_e$ & Availability of server $e$. \\
$d_e$ & The last task finished by $e$. \\
$\mathbf{G}_m^t$ & Set of servers that load model $m$ at time $t$. \\
$t^w_k$ & Waiting time for task $k$. \\
$t^d_k$ & Model initialization time for task $k$. \\
$t^r_k$ & Inference latency of task $k$. \\
$t^{\text{avg}}_{\mathbf{Q},t}$ & Average waiting time of tasks in the queue at time $t$. \\
$p_k$ & Time penalty for running task $k$. \\
$p_{quality}$ & Penalty value for substandard quality. \\
$q_k$ & Quality score of task $k$ using CLIP model. \\
$q_{min}$ & Minimum quality threshold required by the system. \\
$I_k$  & Quality penalty term for task $k$. \\
$U_k$ & Utility of running task $k$. \\
$\mathcal{D}_{g}$ & Probability distribution for task generation interval. \\
$\mathcal{D}_{c}$ & Probability distribution for task collaboration server. \\
$\alpha_q$ & Quality weight coefficient in optimization objective. \\
$\beta_t$ & Time weight coefficient of response time. \\
$\mu_t$ & Penalty weight coefficient of average queue waiting time. \\
$\lambda_q$ & Penalty weight coefficient in optimization objective. \\
\hline
\end{tabular}
\label{tab:notations}
\end{table}
\fi


\subsection{System Model}

In our EAT system, users submit AIGC tasks for processing. Each task includes a prompt and is assigned a number of GPUs. Upon submission, tasks are stored sequentially in a queue for execution. Each time slot, the scheduler evaluates the top $l$ tasks in the queue, along with the servers' running models and load information, to select tasks for execution, determine the inference steps, and assign them to the appropriate servers. Once a task is completed, the resulting image is returned and evaluated using Contrastive Language-Image Pretraining (CLIP) scores \cite{b4.4}. The main components are as follows.


\subsubsection{Task} We consider a set of tasks $\mathbf{K} = \{k_1, k_2, \dots\}$ submitted by users, ordered by arrival time. Each task $k \in \mathbf{K}$ is defined by a tuple: $k = (g_k, c_k, t^a_k)$ where $g_k$ is the text prompt for content generation, $c_k$ is the number of parallel servers required for execution (a random variable sampled from the discrete distribution $\mathcal{D}_{c}$), and $t^a_k$ is the arrival timestamp.

The system exhibits dual randomness in task characteristics. First, the collaboration requirement $c_k \sim \mathcal{D}_{c}$ with $c_k \in \{1, 2, 4, 8\}$ reflects varying task scale. Second, the generation interval time $t^g_k = t^a_{k+1} - t^a_k$ between consecutive tasks follows $t^g_k \sim \mathcal{D}_{g}$, where $\mathcal{D}_{g}$ is a random distribution suitable for modeling task arrival patterns. The arrival time follows the recursive relation $t^a_{k+1} = t^a_k + t^g_k$ for $k \geq 1$.

All arriving but unscheduled tasks are stored in a waiting queue $\mathbf{Q}(t)$ ordered by arrival time. Each task exhibits execution characteristics including execution time $t^e_k = f(s_k, c_k)$ related to inference steps and parallelism \cite{b2.3}, initialization time $t^d_k = g(c_k, m_k)$ for model loading, and generation quality $q_k = h(s_k, g_k)$ determined by inference steps.


\subsubsection{Edge System} The edge computing environment consists of a server cluster $\mathbf{E} = \{e_1, e_2, \dots, e_{|\mathbf{E}|}\}$. Each server $e \in \mathbf{E}$ at time $t$ is characterized by $\{a_e(t), t^r_e(t), d_e(t)\}$
where $a_e(t) \in \{0, 1\}$ indicates availability (1 for idle, 0 for busy), $t^r_e(t) \geq 0$ represents the estimated remaining completion time of the current task and $d_e(t)$ denotes the loaded model type (0 indicates no model).

For model reuse, we define the set of idle servers with the same model type as:
\begin{equation}
\mathbf{G}_m^t = \{e \in \mathbf{E} \mid a_e(t) = 1 \land d_e(t) = m\}.
\end{equation}
When scheduling task $k$ requiring $c_k$ servers and $|\mathbf{G}_{m_k}(t)| = c_k$, model reuse can be achieved.

\subsubsection{Scheduler} The scheduler makes composite decisions at each decision moment $t$. The decision space contains three parts: $a^{exec}_t \in \{0, 1\}$ determines whether to start a new task at the current moment. $a^{task}_t \in \{1, 2, \dots, |\mathbf{Q}(t)|\}$ selects a task from the waiting queue for execution. $a^{steps}_t \in [S_{min}, S_{max}]$ determines the Stable Diffusion inference steps for the selected task to balance time-quality trade-offs.

The scheduler must satisfy the following constraints: at least $c_k$ idle servers ($|\{e \in \mathbf{E} : a_e(t) = 1\}| \geq c_k$); inference steps within the range $S_{min} \leq a^{steps}_t \leq S_{max}$; and the Gang scheduling condition, requiring a subset $\mathbf{E}_k \subset \mathbf{E}$ with $|\mathbf{E}_k| = c_k$ where every $e \in \mathbf{E}_k$ is idle ($a_e(t) = 1$).

\subsubsection{Task Metrics} System performance is quantified through time cost and generation quality dimensions.

\textbf{Response Time}: Task $k$'s response time (total time from submission to completion) depends on model reuse availability. If a server group $\mathbf{G}_{k'}$ with $|\mathbf{G}_{k'}| = c_k$ exists, the AIGC model is already loaded, enabling reuse without initialization overhead: $t^r_k = t^e_k + t^w_k$. Otherwise, additional initialization delay $t^d_k$ is required: $t^r_k = t^e_k + t^d_k + t^w_k$, where $t^w_k = t^s_k - t^a_k$ represents waiting time.

\textbf{Quality Score}: Task $k$'s quality is evaluated using the CLIP model. For generated image $i_k$ and corresponding text prompt $g_k$, the quality score is calculated as:
\begin{equation}
    q_k = w_q \times \text{CLIP}(i_k, g_k),
\end{equation}
where $w_q$ is the quality weight. Higher scores indicate stronger semantic matching between generated images and prompts.

\textbf{Quality Penalty}: To ensure output quality standards, we introduce a quality penalty term $I_k$ defined as:
\begin{equation}
    I_k = \begin{cases}
    p_{quality} & \text{if } q_k < q_{min} \\
    0 & \text{if } q_k \geq q_{min}
    \end{cases},
\end{equation}
where $q_{min}$ is the minimum quality threshold and $p_{quality}$ is the penalty value for substandard quality.

To minimize task inference latency, we define the utility of task \(k\) as \(U_k = q_k - p_k\) and aim to maximize the total global utility.

\subsection{Optimization Objective}
The scheduler aims to enhance generation quality and minimize response time while maintaining a minimum quality standard. We define the optimization objective as a utility function that balances quality and response time.

\newtheorem{problem}{Problem}
\begin{problem} 
\begin{subequations} \label{eq:optimization_problem}
\begin{align}
\max_{\pi} \quad & \mathcal{U} = \mathbb{E}_{\pi} \left[ \sum_{k \in \mathbf{K}} (\alpha_q q_k - \beta_t t^r_k - \lambda_q I_k) \right] \label{eq:objective} \\
\text{s.t.} \quad 
%
& |\{e \in \mathbf{E} : a_e(t) = 1\}| \geq c_k, \text{when scheduling } k \label{eq:constraint_availability} \\
&  \forall k \in \mathbf{K},\forall e \in \mathbf{E}_k, |\mathbf{E}_k| = c_k \text{ and } a_e(t) = 1  \label{eq:constraint_gang} \\
&  \forall k \in \mathbf{K}, S_{\min} \leq s_k \leq S_{\max},  \label{eq:constraint_steps} 
\end{align}
\end{subequations}
\end{problem}
where $\alpha_q$, $\beta_t$, and $\lambda_q$ are the quality, time, and penalty weight coefficients. The objective function maximizes generation quality, minimizes response latency, and penalizes low-quality outputs. Constraints ensure quality assurance, resource availability, gang scheduling, and task generation. Traditional methods cannot effectively address these complexities. Modeling the problem as an MDP enables the use of RL to solve the EAT problem.

\section{Algorithm Design} \label{Algorithm Design}

\subsection{MDP Formulation}

We model the AIGC task scheduling problem as a continuous-time, discrete-decision MDP represented by the tuple $(\mathcal{S}, \mathcal{A}, P, R, \gamma)$.

\subsubsection{State $\mathcal{S}$}

The system state $S_t \in \mathcal{S}$ is a high-dimensional vector:
\begin{equation}
    S_t = \left(\mathbf{E}_{\text{state}}(t), \mathbf{Q}_{\text{state}}(t)\right),
\end{equation}
where the cluster state $\mathbf{E}_{\text{state}}(t) = \{a_e(t), t^r_e(t), d_e(t) \mid \forall e \in \mathbf{E}\}$ captures server availability, remaining time, and loaded model types. The queue state $\mathbf{Q}_{\text{state}}(t) = \{(c_k, t^a_k) \mid \forall k \in \mathbf{Q}(t)\}$ represents top $l$ tasks collaboration requirements and waiting times.

To unify the server and task queue states, we concatenate the server state vector with the task state vector. The server state is a \(3 \times \left |  \mathbf{E} \right | \) matrix, while the task state, initially a \(2 \times l\) matrix, is expanded to a \(3 \times l\) matrix by adding a row of zeros. Therefore, the final state \(s\) is a \(3 \times (\left |  \mathbf{E} \right |+l)\) matrix, defined as follows:
\begin{equation}
s = \underbrace{
\begin{bmatrix}
{a}_{e_1} & \cdots & {a}_{e_{\left | \mathbf{E} \right |} } & t^a_{k_1} & \cdots & t^a_{k_{\left | \mathbf{K} \right |}} \\
t^r_{e_1}  & \cdots & t^r_{e_{\left |\mathbf{E}  \right |} }  & c_{k_1} & \cdots & c_{k_{\left | \mathbf{K} \right |} }\\
d_{e_1}    & \cdots & d_{e_{\left | \mathbf{E} \right |} }     & 0   & \cdots & 0
\end{bmatrix}
}_{3 \times ({\left |  \mathbf{E} \right |}+l)}.
\end{equation}


\subsubsection{Action $\mathcal{A}$}

The action space \(\mathcal{A}\) represents scheduling decisions at each time step, which includes three parts:
\begin{equation}
    \mathcal{A} = \mathcal{A}^{exec} \times \mathcal{A}^{task} \times \mathcal{A}^{steps},
\end{equation}
where $\mathcal{A}^{exec} = \{0, 1\}$ determines execution decisions, $\mathcal{A}^{task} = \{1, 2, \dots, |\mathbf{Q}(t)|\}$ selects tasks from the queue, and $\mathcal{A}^{steps} = [S_{min}, S_{max}]$ determines inference steps. The total action complexity is $2 \times |\mathbf{Q}(t)| \times (S_{max} - S_{min} + 1)$.

To simplify the action space, we transform the discrete action components into a continuous variable sequence. The action \(a\) can then be formalized as:
\begin{equation}
a^\top = \begin{bmatrix}
a_c & a_s & a_{k_1} & a_{k_2} & \cdots & a_{k_l}
\end{bmatrix}.
\end{equation}

The action vector comprises three components that guide scheduling decisions. The execution decision \(a_c \in [0,1]\) determines whether to schedule a task at the current time step: \(a_c \leq 0.5\) means a task is scheduled, while \(a_c > 0.5\) means no action is taken. The inference step component \(a_s \in [0,1]\) indicates the diffusion steps for the chosen task, mapped linearly to the discrete range \([S_{min}, S_{max}]\). The task selection components \([a_{k_1}, a_{k_2}, \dots, a_{k_l}]\) assign preference scores (\(a_{k_i} \in [0,1]\)) to each task in the queue. When scheduling occurs, the task with the highest preference is selected.

\subsubsection{Transition Function $P$} The state transition $P(S_{t+1} | S_t, A_t)$ encompasses two mechanisms:

\textbf{Task Queue Transitions}: New tasks arrive according to intervals $t^g_k \sim \mathcal{D}_{g}$ and collaboration requirements $c_{k+1} \sim \mathcal{D}_{c}$. When $t = t^a_{k+1} = t^a_k + t^g_k$, task $k+1$ joins the queue: $\mathbf{Q}(t+1) = \mathbf{Q}(t) \cup \{k+1\}$. Scheduled tasks are removed: $\mathbf{Q}(t+1) = \mathbf{Q}(t) \setminus \{k^*\}$ when action $A_t$ selects task $k^*$.

\textbf{Server State Transitions}: When task $k^*$ is scheduled to server set $\mathbf{E}_{k^*}$, server states update as: $a_e(t+1) = 0$, $t^r_e(t+1) = t^e_{k^*} + t^d_{k^*}$ (with initialization) or $t^r_e(t+1) = t^e_{k^*}$ (model reuse), and $d_e(t+1) = k^*$ for $e \in \mathbf{E}_{k^*}$. For running servers, remaining time decreases: $t^r_e(t+1) = \max(0, t^r_e(t) - \Delta t)$. The remaining time $t^r_e$ is predicted based on the characteristics of AIGC tasks, akin to \cite{b2.3}. We experimentally measure the initialization overhead of the AIGC service, specifically the Stable Diffusion model with various patches, as shown in Table~\ref{tab:time_predict}. The results indicate that initialization time remains relatively constant. We measure the time per inference step during task execution, as shown in Table~\ref{tab:time_predict}, which reveals a linear relationship with the number of inference steps.

\begin{table}[htbp]
\centering
\caption{Time Prediction}
\begin{tabular}{c|c|c}
\hline
\textbf{Patch Number} & \textbf{Init Time (s)}& \textbf{Time per Inference Step (s)}\\
\hline
1& 33.5& 0.53\\
2& 31.9& 0.29\\
4& 35.0& 0.20\\
\hline
\end{tabular}
\label{tab:time_predict}
\end{table}

\subsubsection{Reward $R$} To align with the optimization goal and discourage excessive task delays, we define the immediate reward as:
$$R_t = \alpha_q \cdot q_{k^*} - \lambda_q \cdot I_{k^*} + \frac{1}{\beta_t \cdot t^r_{k^*} + \mu_t \cdot t^{\text{avg}}_{\mathbf{Q},t}}$$
where $k^*$ is the task scheduled at time $t$. $\alpha$, $\beta$, and $\lambda$ are quality, time, and penalty coefficients consistent with the optimization objective, respectively. $\mu_t$ is the queue penalty coefficient, and $t^{\text{avg}}_{\mathbf{Q},t} = \frac{\sum_{k \in \mathbf{Q}} (t - t^a_k)}{|\mathbf{Q}|}$ is the average waiting time in the queue. To prevent extremely delayed tasks from causing extreme reward values, we employ a reciprocal form for time penalty rather than direct subtraction.

\subsection{Proposed EAT Algorithm}

\begin{figure}[t]
    \centering
    \includegraphics[width=0.92\columnwidth, page=1]{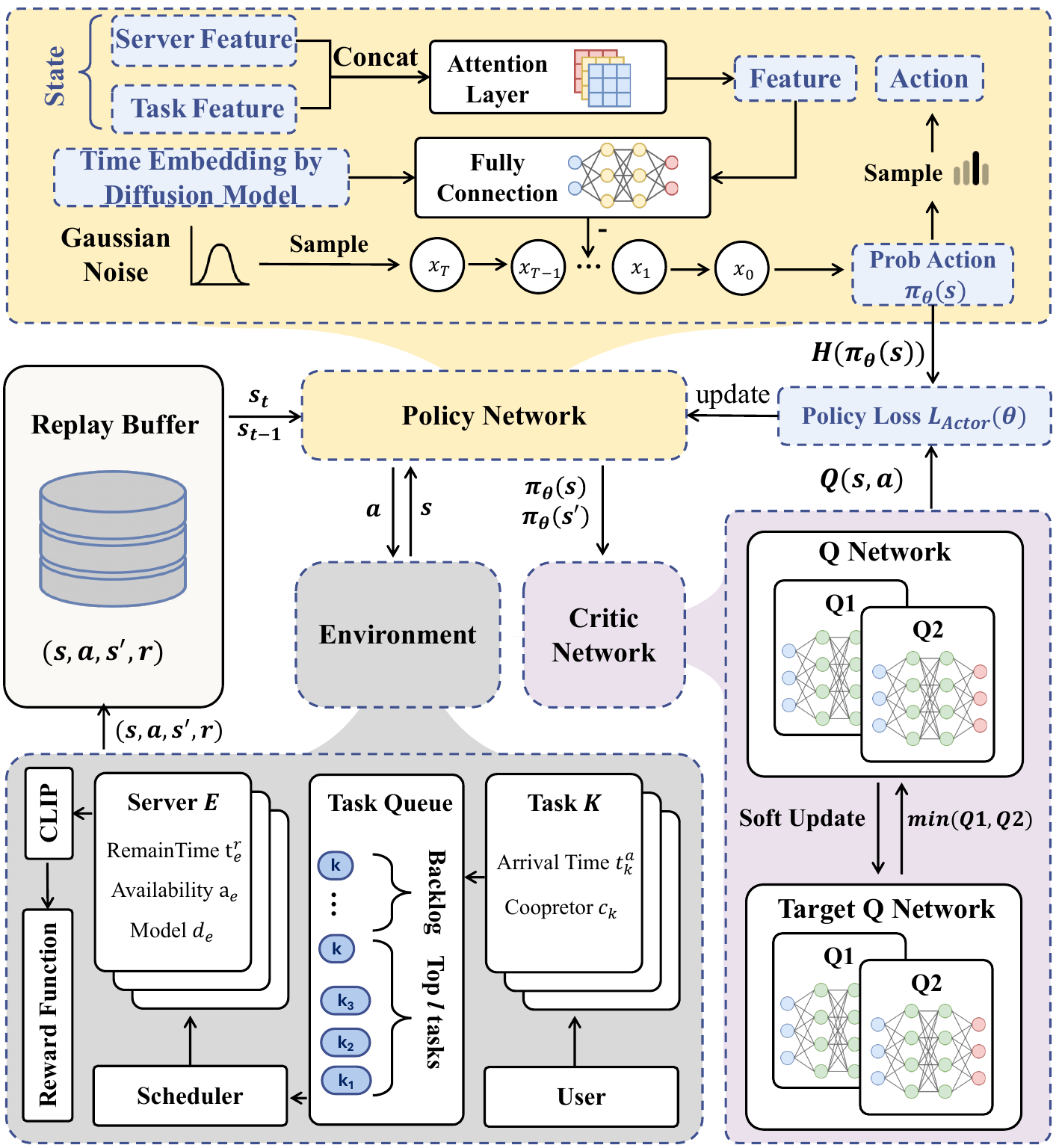} 
    \caption{EAT Algorithm Framework.}
    \label{Framework}
\end{figure}

The decision process of EAT, as outlined in Algorithm \ref{algorithm:decision}, begins by initializing the environment to obtain the initial state $s$. The EAT algorithm employs an Attention layer to extract features $f_s$ from the state, which are then fed into the diffusion model to generate an action $a_t$ (line 4). The Task Selector uses $a_t$ to choose task $k$ from the queue, while the Server Selector picks server $\mathbf{E}_k$ to execute it, as illustrated in lines 5-6. If the server in $\mathbf{E}_k$ is unavailable or the EAT algorithm chooses not to execute the task, no decision is made (lines 7-9). Otherwise, task $k$ is deployed into the environment based on $a_t$ and $\mathbf{E}_k$. The execution time is predicted based on the task's inference steps, the environment is updated, and the next state $s'$ is determined (lines 11-13). This continues until all tasks are finished or a time limit is reached. The framework of EAT algorithm is shown in Fig~\ref{Framework}.

\begin{algorithm}[t]
\begin{algorithmic}[1]
\caption{Decision Process of EAT Algorithm}
\label{algorithm:decision}
\Require Trained policy network \(\pi_\theta\)
\State Load EAT agent with policy \(\pi_\theta\)
\State Reset environment and observe initial state \(s\)
\While{not done}
    \State Use Attention layer to get $f_s$
    \State Initialize a random distribution $x_T \sim \mathcal{N}(0, I)$
    \For{$t = T$ to $1$}
        \State Predict a denoising distribution $\epsilon_\theta(x_t, t, f_s)$
        \State Compute the reverse transition distribution $p_\theta(x_{t-1} \mid x_t)$ using Eqs. (\ref{diffusion p}) and (\ref{diffusion mu})
        \State Sample $x_{t-1}$ from $p_\theta(x_{t-1} \mid x_t)$
    \EndFor
    \State Compute the distribution of $x_0$ using Eq. (\ref{diffusion x_0})
    \State Sample action $a \sim x_0$
    \If{not excute according to \(a_c\)}
        \State Observe new state \(s'\)
        \State \textbf{Continue}
    \EndIf
    \State Compute task probabilities \(prob_k = \text{Softmax}(a_k)\)
    \State Select task \(k = \arg\max(prob_k)\)
    \State Determine reasoning steps \(s_k\) using \(a_s\)
    \If{existing model group \(|\mathbf{G}_m^t| = c_k\)}
        \State Assign edge server(s) \(E_k = \mathbf{G}_m^t\)
    \Else
        \If{existing sufficient idle edge nodes}
            \State Unload old model(s) from selected node(s)
            \State Load model in \(c_k\) nodes
        \Else
            \State Observe new state \(s'\)
            \State \textbf{Continue}
        \EndIf
    \EndIf
    \State Execute task $k$ on the edge server(s)
    \State Predict task excution time $t^e_{k}$ 
    \State Observe next state \(s'\)
\EndWhile
\end{algorithmic}
\end{algorithm}

\subsubsection{Feature Extraction}

The attention mechanism minimizes information overload and captures global dependencies by weighting inputs, allowing the model to emphasize key features \cite{b4.3}. The attention score $\text{Attention}(Q, K, V)$ is calculated using a set of key vectors \(K\), a query vector \(Q\), and a set of value vectors \(V\):
\begin{equation}
\begin{matrix}
\text{Attention}(Q, K, V) = \text{softmax}\left(\frac{QK^\top}{\sqrt{d_K}}\right) V,
\end{matrix}
\end{equation}
where \(d_K\) is a scaling factor corresponding to the dimension of the key vectors. We utilize this attention mechanism to efficiently extract features from the state sequence, as demonstrated in line 4 of Algorithm \ref{algorithm:decision}. Each column of the state matrix is treated as a vector, resulting in a sequence of vectors. This sequence is processed in an attention layer, producing a feature vector $f_s$. This compact representation enhances decision-making in the scheduling process.

\subsubsection{Diffusion-based Policy}

\begin{figure*}[!t]
    \centering
    \includegraphics[width=0.87\linewidth, page=1]{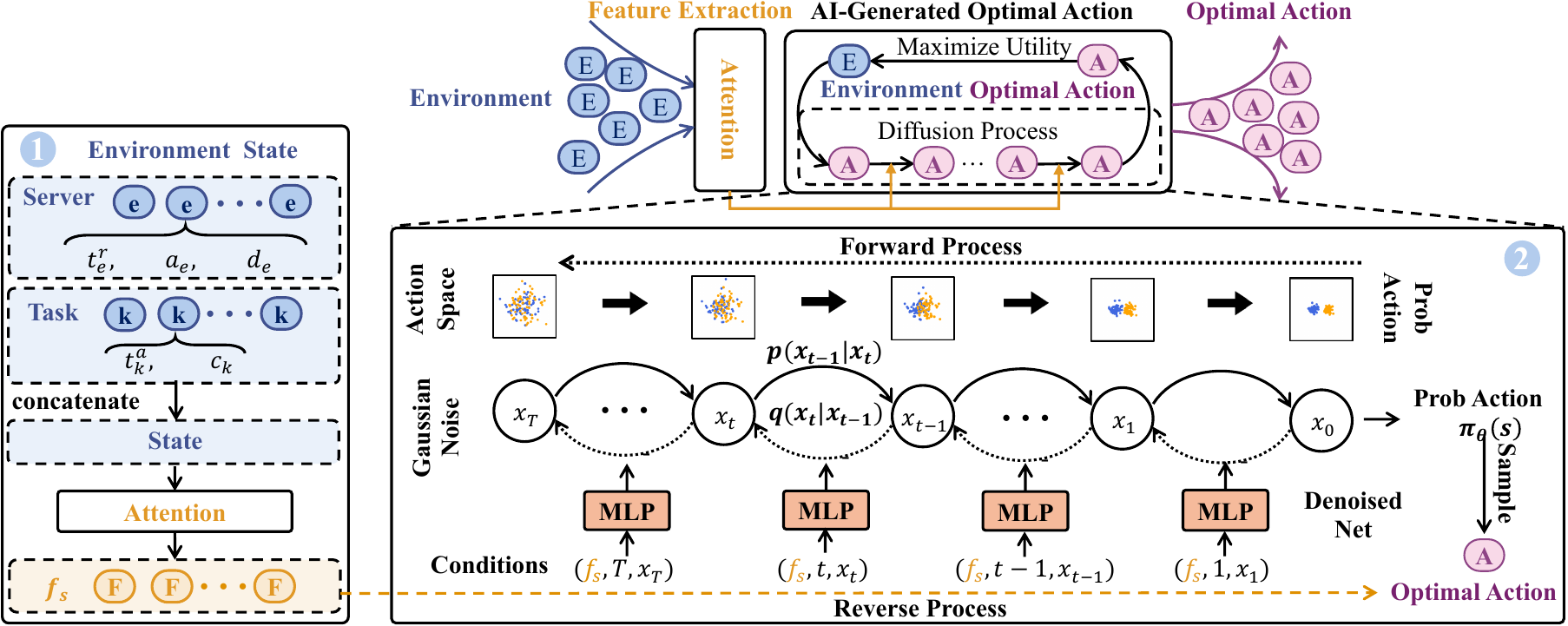} 
    \caption{Diffusion-based Policy in EAT.}
    \label{algorithm}
\end{figure*}

A policy \(\pi\) defines a rule for action selection. As illustrated in Fig~\ref{algorithm}, we combine the diffusion model with the policy network \cite{b2.1}. The diffusion model adds noise to the data until it becomes Gaussian, then reverses the process by starting with Gaussian noise and progressively restoring the original data using an inverse conditional distribution. A learned inverse model functions as a denoiser to reconstruct the original action from the noisy data. The reverse diffusion is defined as:
\begin{equation}
\begin{matrix}
p_\theta(x_{i-1} \mid x_i) = \mathcal{N}\Bigl(x_{i-1}\mid \mu_\theta(x_i, i),\; \beta_i I \cdot \frac{1-\bar{a}_{i-1}}{1-\bar{a}_i}\Bigr),\label{diffusion p}
\end{matrix}
\end{equation}
where \(\beta_i\) is a deterministic variance amplitude that can be easily computed \cite{b4.1}. Using Bayes’ formula, the reverse process is reformulated into a Gaussian probability density function to compute the mean. Each denoising step \(i\) in the reverse process introduces independent noise \(\epsilon_\theta\), distinct from the forward process noise \(\epsilon\). Consequently, the denoised output \(x_0\) is computed as:
\begin{equation}
\begin{matrix}
x_0 = \frac{1}{\sqrt{\bar{a}_i}}\, x_i - \frac{1}{\sqrt{\bar{a}_{i-1}}} \cdot \tanh\Bigl(\epsilon_\theta(x_i, i, f_s)\Bigr), \label{diffusion x_0}
\end{matrix}
\end{equation}
and the mean is estimated by:
\begin{equation}
\begin{matrix}
\mu_\theta(x_i, i, f_s) = \frac{1}{\sqrt{a_i}}\, x_i - \beta_i\, \frac{\epsilon_\theta(x_i, i, f_s)}{\sqrt{1-\bar{a}_i}}, \label{diffusion mu}
\end{matrix}
\end{equation}
for \(i = 1, \ldots, T\), where \(\epsilon_\theta(x_i, i, f_s)\) is produced by a deep model with parameters \(\theta\) conditioned on the feature \(f_s\).

In our EAT algorithm, the continuous action vector leads the reverse diffusion process to produce a mean \(x_0\) for the action. According to the Soft Actor Critic (SAC) \cite{b4.2} algorithm, we feed the mean \(x_0\) into an additional linear layer to generate a variance vector \(\sigma^2(x_0)\) of the same dimensionality. The final continuous action is sampled from a Gaussian distribution with mean \(x_0\) and variance \(\sigma^2(x_0)\):
\begin{equation}
a_j \sim \mathcal{N}\Bigl(x_{0,j},\; \sigma^2_{\theta,j}(x_0)\Bigr), \quad \forall\, j.
\end{equation}
This approach separates the action mean from the action variance, enabling more stable and efficient policy training.

\subsubsection{Task Selector}

The actor chooses a task from the queue and determines the necessary diffusion steps for optimal server workload balancing. At each decision step, the actor outputs $a$, with each element serving a specific purpose. The task selector evaluates the action component $\begin{bmatrix} a_{k_1} & a_{k_2} & \cdots & a_{k_l} \end{bmatrix}$ as task scores and chooses the task with the highest score for execution. The scheduler then retrieves the selected task from the queue and prepares it for execution, as illustrated in line 5 of Algorithm \ref{algorithm:decision}.

\subsubsection{Server Selector}

Once the task execution parameters are determined, an appropriate server must be selected for execution, as outlined in line 6 of Algorithm \ref{algorithm:decision}. We use a greedy strategy for server selection. Specifically:
\begin{itemize}
    \item At time $t$, if an idle group \(\mathbf{G}_m^t\) has available servers such that \(|\mathbf{G}_m^t| = c_k\), then that group is selected.
    \item The strategy chooses servers that minimize idle group fragmentation. It first terminates the processes linked to those servers, then deploys the new AIGC model.
\end{itemize}

\begin{algorithm}[t]
\begin{algorithmic}[1]
\caption{Training of EAT Algorithm}
\State Initialize parameters \(\theta\), \(\phi_1\), \(\phi_2\), \(\phi_1' \gets \phi_1\), \(\phi_2' \gets \phi_2\)
\State Initialize experience replay buffer $\mathbf{D}$
\For{$episode = 1, 2 \dots$}
    \State Initialize environment and get state $s_0$
    \While{not done}
        \State Use Attention layer to get $f_s$
        \State Initialize a random distribution $x_T \sim \mathcal{N}(0, I)$
        \For{$t = T$ to $1$}
            \State Scale the inferred distribution $\epsilon_\theta(x_t, t, f_s)$
            \State Compute the mean $\mu_\theta$ of the reverse transition distribution $p_\theta(x_{t-1} \mid x_t)$ using Eqs. (\ref{diffusion p}) and (\ref{diffusion mu})
            \State Sample $x_{t-1}$ from $p_\theta(x_{t-1} \mid x_t)$
        \EndFor
        \State Compute the distribution of $x_0$ using Eq. (\ref{diffusion x_0})
        \State Sample action $a \sim x_0$
        \State Execute action $a$ in environment
        \State Get the next state $s'$ and reward $r$
        \State Store the transition $(s, a, s', r)$ into the buffer $\mathbf{D}$
    \EndWhile
    \State Randomly sample a batch of experience from $\mathbf{D}$
    \State Update policy network parameters $\theta$ using Eqs.~(\ref{actor loss})  and (\ref{actor update})
    \State Update critic network parameters $\phi_1$, $\phi_2$ using Eq.~(\ref{critic update})
    \State Update parameters $\hat{\phi}_1$, $\hat{\phi}_2'$ using Eq. (\ref{cirtic soft update})
\EndFor
\State Return the optimized policy network $\pi_\theta$, with trained parameters $\theta$
\label{training}
\end{algorithmic}
\end{algorithm}

\subsection{Training}

In the training process shown in Algorithm~\ref{training}, the policy network parameters $\theta$, critic network parameters $\phi_1$, $\phi_2$, target network parameters $\phi_1'$ , $\phi_2'$, and the experience replay buffer $\mathbf{D}$ are initialized as shown in lines 1-2. In each episode, the environment is first initialized, and the initial state $s_0$ is obtained as shown in line 4. The interaction with the environment then begins, continuing until all tasks are completed or the time limit is reached. During each interaction, the attention layer is used to extract features $f_s$ from the state, and the diffusion model generates an action, as shown in lines 6-15. The action is then executed in the environment, resulting in the next state $s'$ and reward $r$. The experience $(s, a, s', r)$ is stored in the experience replay buffer $\mathbf{D}$ as shown in lines 15-17. After an episode's experience is collected, a batch of experiences is sampled from the replay buffer to update the actor network, update the critic network, and perform soft updates to the target critic networks, as shown in lines 19-22. After training is completed, the policy parameters $\theta$ are obtained.

\subsubsection{Actor Training}
To prevent the policy from becoming overly confident in certain actions and converging to suboptimal solutions, we add an action entropy regularization term to encourage exploration \cite{b4.2}. Our approach models the action as a continuous Gaussian distribution, parameterized by a mean and variance. The mean is derived from the diffusion process, while the variance is obtained by passing the mean through an additional linear layer, ensuring both share the same dimensionality as the action vector.

The entropy \(H\) of a Gaussian distribution with diagonal covariance is defined as:
\begin{equation}
\begin{matrix}
H\Bigl(\mathcal{N}(\mu, \sigma^2)\Bigr) = \frac{1}{2} \sum_{j=1}^{d} \log\bigl(2\pi e\, \sigma_j^2\bigr),
\end{matrix}
\end{equation}
where \(d\) is the dimensionality of the action space. Incorporating the entropy term, the policy objective is formulated as:
\begin{equation}
\begin{aligned}
J(\theta) = \mathbb{E}_{s_t \sim B_e} \Bigl[ Q_{\phi}^e(s_t, a_t) +\alpha\, H\bigl(\mathcal{N}(\mu_\theta(s_t), \sigma^2_\theta(s_t))\bigr) \Bigr],
\end{aligned}
\end{equation}
where \(\alpha\) is the temperature coefficient that controls the strength of entropy regularization. The gradient of the actor loss is then:
\begin{equation}
\begin{aligned}
\nabla_\theta L_{\theta}^e = \mathbb{E}_{s_t \sim B_e} \Bigl[ &-\alpha\, \nabla_\theta H\Bigl(\mathcal{N}(\mu_\theta(s_t), \sigma^2_\theta(s_t))\Bigr)\\ &- \nabla_\theta \, Q_{\phi}^e(s_t, a_t) \Bigr].
\label{actor loss}
\end{aligned}
\end{equation}
The actor network is updated with learning rate \(\eta_a\):
\begin{equation}
\theta^{e+1} \leftarrow \theta^e - \eta_a\, \nabla_\theta L_{\theta}^e.
\label{actor update}
\end{equation}

\subsubsection{Critic Network}
To reduce overestimation bias in value estimation, the EAT algorithm utilizes double critic networks \cite{b4.2}. Each critic network has its parameters, \(\phi_1\) and \(\phi_2\). To enhance training stability, corresponding target networks are maintained for calculating the target Q value and updating the Q network, with parameters \(\phi_1'\) and \(\phi_2'\). The actor optimizes based on the minimum Q-value from the two critic networks:
\begin{equation}
Q_\phi(s_t,a_t) = \min \left\{ Q_{\phi_1}(s_t,a_t),\; Q_{\phi_2}(s_t,a_t) \right\}.
\end{equation}

\subsubsection{Critic Training}
The Q-function loss is defined as:
\begin{equation}
L_{\phi_i}^e = \mathbb{E}_{(s_t, a_t, s_{t+1}, r_t) \sim B_e} \left[ \left(\hat{y}^e - Q_{\phi_i}^e(s_t, a_t)\right)^2 \right], \ i = 1, 2
\label{cirtic loss}
\end{equation}
where the target value \(\hat{y}^e\) is computed as:
\begin{equation}
\hat{y}^e = r_t + \gamma\hat{Q}_{\phi}^e(s_{t+1},a_{t+1}),
\end{equation}
where \(\gamma\) is the discount factor for future rewards.

The critic networks are updated by performing one step of gradient descent to minimize the critic loss \(L_{\phi_i}^e\). 
\begin{equation}
\phi_i^{e+1} \leftarrow \phi_i^e - \alpha \nabla_{\phi_i} L_{\phi_i}^e
\label{critic update}
\end{equation}
In addition, to stabilize learning, the critic's target network parameters \(\hat{\phi}\) are updated using a soft update rule:
\begin{equation}
\hat{\phi}^{e+1} \leftarrow \tau\, \phi^{e+1} + (1-\tau)\, \hat{\phi}^e
\label{cirtic soft update}
\end{equation}
where \(\tau \in (0,1]\) is the target update rate. Note that in our framework, only the critic network employs a target network; the actor network is updated directly using its loss.

\subsection{Computational Complexity Analysis}

For the training process, we perform a total of $S$ environment steps, each involving the following operations. 1) Feature Extraction. The forward pass complexity is $O((|\mathbf{E}|+l)^2 \cdot d)$, where $d$ is the dimension of the state. 2) Diffusion Process. The diffusion strategy involves $T$ denoising steps to generate actions, with each step having a complexity of $f_\theta$, leading to a total complexity of $O(T \cdot f_\theta)$. 3) Task and Server Selection. The complexity is $O(|\mathbf{E}| + l)$. Therefore, the total complexity over $S$ environment steps is
$O(S \cdot ((|\mathbf{E}|+l)^2 \cdot d + T \cdot f_\theta + (|\mathbf{E}|+l)))$.

At the end of each episode, the algorithm updates parameters, including the policy network and two critic networks. With $E$ episodes in total, the complexity of parameter updates becomes
$O(E \cdot (|\theta| + 2|\phi|))$. Combining the above, the total computational complexity of the algorithm is given by $O\left( S \cdot \left((|\mathbf{E}|+l)^2 \cdot d + T \cdot f_\theta + (|\mathbf{E}|+l)\right) + E \cdot (|\theta| + 2|\phi|) \right)$.

\section{Experiment And Result} \label{Experiment And Result}

\begin{figure*}[!t] \centering \includegraphics[width=0.96\linewidth]{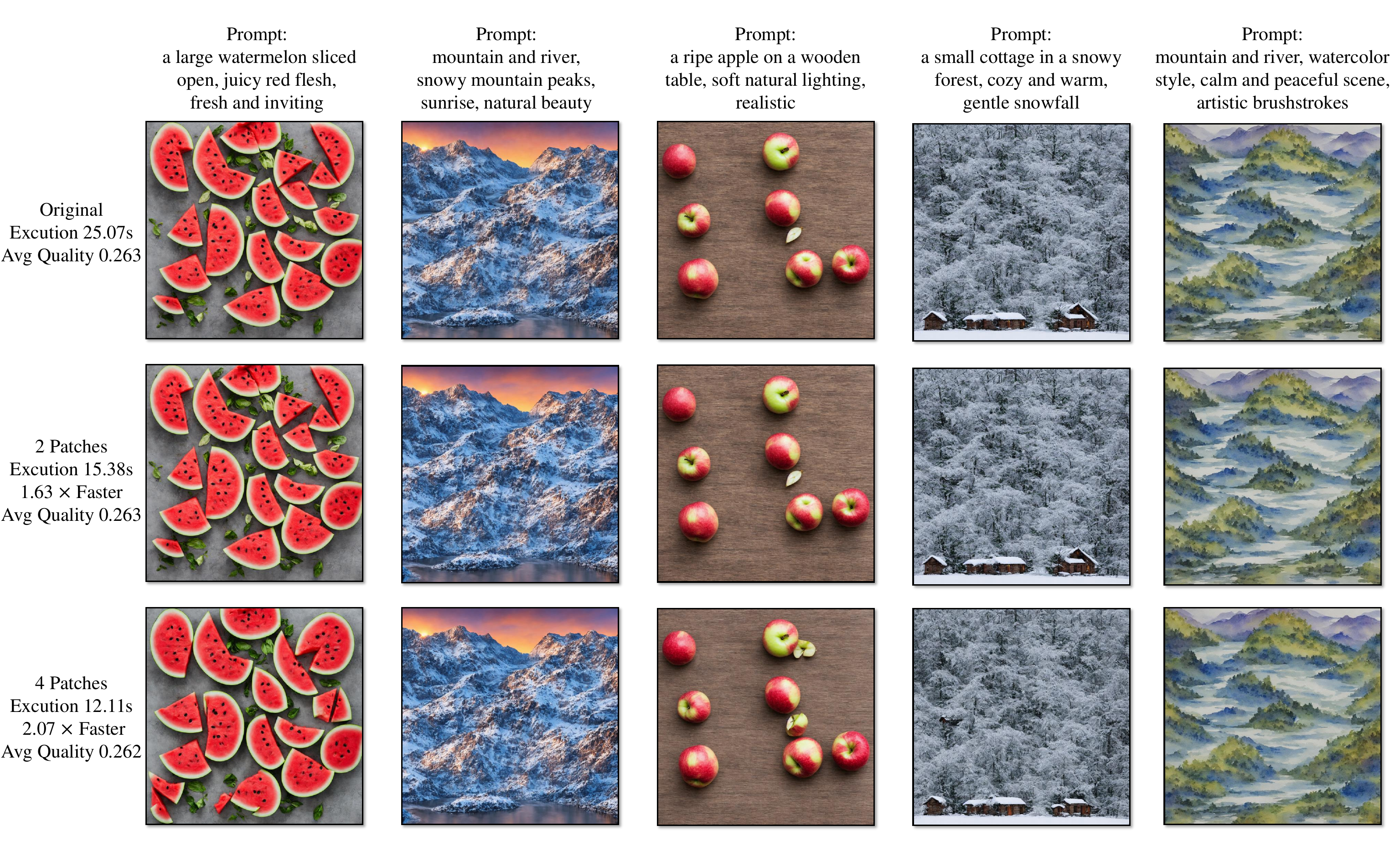} \caption{The image results and excution time of our edge computing system implementation using Stable Diffusion.} \label{demo} \end{figure*}

\subsection{Experimental Settings}

\subsubsection{System Implementation}

For model training, we utilize a workstation equipped with an Intel Core i9-14900KF CPU, 64GB of RAM, and an NVIDIA GeForce RTX 4090 24G GPU, running Ubuntu 22.04.5 LTS with CUDA 12.6. For validation, we set up an edge cluster featuring Intel Xeon Platinum 8336C CPUs, 128GB of RAM, and four NVIDIA GeForce RTX 4090 24G GPUs, also using Ubuntu 22.04.5 LTS with CUDA 12.4. The validation phase deploys four independent Docker containers, each assigned one GPU to perform distributed inference tasks using the DistriFusion framework \cite{b3.4} integrated with RL. We utilize NVIDIA Collective Communications Library(NCCL) 2.21.5 for efficient GPU communication. Torch dynamically creates process groups encompassing different GPUs for each distributed task, enabling inter-node communication via NCCL using distinct IPs and ports to emulate a multi-node edge environment.

Each container acts as a server, using a socket to listen on a designated port for commands from the host. When assigned a task, the host packages the task details into a JSON string and sends it via the socket to the server responsible for execution. Each JSON string contains the user-submitted task prompt $p_k$ and the draw step  $s_k$ determined by the algorithm. After dispatching the command, the host asynchronously monitors the server's result port. Once the task is completed, the server asynchronously sends the resulting image and JSON back through the result port, and the host receives it, marking the task as complete. The host then retrieves detailed statistics from the result JSON, such as the actual task execution time and actual model loading time.

We primarily conduct experiments using the popular public text-to-image model, Stable Diffusion v1.4\cite{b3.2}. The model first compresses an image into an 8 $\times$ smaller latent representation through a pre-trained autoencoder, and then applies a denoising diffusion process in the latent space. To guide the generation process, Stable Diffusion uses CLIP to encode text prompts into semantic vector embeddings as a conditional information to steer denoising process. By default, we use the Denoising Diffusion Implicit Models(DDIM) \cite{b3.2} as sampler with a classifier-free guidance scale of 5, which allows us to generate high-quality images at a resolution of 1440x1440 pixels.

\subsubsection{Parameter Settings}

Each training episode is configured with a time limit of 1024 seconds and a maximum of 1024 decision steps, containing 32 tasks per episode. The model undergo $1.5\times 10^6$ training episodes using the Adam optimizer \cite{b4.6}. Task arrival rates are set to 0.05, 0.1, and 0.15 for 4-server, 8-server, and 12-server configurations, respectively, to match varying server processing capacities.

The EAT algorithm uses an attention mechanism to process initial state features, producing a feature vector \(f_s\) of dimension \(\left | \mathbf{Q} \right | + l\). This vector is input into a denoising network with \(256 \times 256\) fully-connected layers for feature reconstruction. The network incorporates timestep encoding as a 16-dimensional vector and environmental states. The critic network consists of two fully-connected layers with 256 nodes. Other RL algorithms use \(256 \times 256\) Multilayer Perceptron (MLP) architectures for actor-critic modules, with D2SAC incorporating a diffusion-inspired denoising network, and PPO servering as on-policy methods. Mish activation functions \cite{b4.5} are applied to all hidden layers, while the policy network's output layer uses tanh to constrain actions to \([-1,1]\). The details are in TABLE~\ref{tab:actor} and TABLE~\ref{tab:hyperparams}.

\begin{table}[t]
    \centering
    \caption{EAT Network Parameters}\label{tab:actor}
    \begin{tabular}{c|c|c|c}
        \hline
        \multicolumn{1}{c|}{\textbf{Networks}} & \textbf{Layer} & \textbf{Activation} & \textbf{Units} \\
        \hline
        \multirow{7}{*}{Actor} & Attention & - & \(\left | \mathbf{E} \right | + l \) \\
        & TimeEmbedding & - & 16 \\
        & Concatenation & - & - \\
        & FullyConnect & Mish & 256 \\
        & FullyConnect & Mish & 256 \\
        & Output & Tanh & $|a|$ \\
        \hline
    \end{tabular}
\end{table}

\begin{table}[t]
    \centering
    \caption{Hyperparameters for DRL Methods}\label{tab:hyperparams}
    \begin{tabular}{c l c}
        \hline
        \textbf{Symbol} & \textbf{Description} & \textbf{Value} \\
        \hline
        $\eta_{\mathrm{a}}$ & Actor network learning rate & $3\times10^{-4}$ \\
        $\eta_{\mathrm{c}}$ & Critic networks learning rate & $3\times10^{-4}$ \\
        $\alpha$ & Entropy regularization temperature & 0.05 \\
        $\tau$ & Soft update weight & 0.005 \\
        $b$ & Batch size & 512 \\
        $\lambda$ & Weight decay coefficient & $1\times10^{-4}$ \\
        $\gamma$ & Reward discount factor & 0.95 \\
        $T$ & Diffusion model denoising steps & 10 \\
        $D$ & Replay buffer capacity & $1\times10^6$ \\
        $E$ & Total training episodes & $5\times10^3$ \\
        $\nu$ & PPO Value function coefficient & 0.5 \\
        $\beta$ & PPO Entropy coefficient & 0.01 \\
        $\epsilon$ & PPO Policy clip threshold & 0.2 \\
        $g$ & PPO Maximum gradient norm & 0.5 \\
        $\lambda_{\text{G}}$ & PPO GAE advantage estimator & 0.95 \\
        \hline
    \end{tabular}
\end{table}

For meta-heuristic baselines, Harmony Search (Harmony) \cite{geem2001new} and Genetic Algorithm (Genetic)\cite{holland1992adaptation} are adopted. Genetic optimizes solutions via evolutionary operators, while Harmony refines solutions through harmony improvisation. Both of them optimize a 2048-steps. For the Genetic Algorithm, we use a population size of 64, 32 generations, and 10 parents. The crossover probability is 1, with a gene mutation probability of 0.1, and 1 elite solution. For the Harmony Search, both the number of improvisations and harmony memory size are 64. The memory consideration and pitch adjustment probabilities are 0.8 and 0.2, respectively, with a pitch adjustment bandwidth of 1.

\begin{table*}
\centering
\caption{Quality}
\label{tb:quality}
\resizebox{\textwidth}{!}{
\begin{tabular}{l|rrrrr|rrrrr|rrrrr}
\multirow{3}{*}{Algorithm} & \multicolumn{15}{c}{Task Arrival Rate}                                                                                                                                      \\ 
\cline{2-16}
                           & \multicolumn{5}{c|}{4 Nodes}                                  & \multicolumn{5}{c|}{8 Nodes}                                  & \multicolumn{5}{c}{12 Nodes}                 \\ 
\cline{2-16}
                           & 0.01   & 0.03   & 0.05   & 0.07   & \multicolumn{1}{r|}{0.09} & 0.06   & 0.08   & 0.1    & 0.12   & \multicolumn{1}{r|}{0.14} & 0.11   & 0.13   & 0.15   & 0.17   & 0.19    \\ 
\hline
{EAT} & {0.256} & {0.264} & {0.262} & {0.262} & {0.262} & {0.260} & {0.261} & {0.261} & {0.261} & {0.262} & {0.264} & {0.262} & {0.260} & {0.260} & {0.262} \\
EAT-A & 0.259 & 0.262 & 0.265 & 0.265 & 0.264 & 0.261 & 0.261 & 0.262 & 0.265 & 0.264 & 0.265 & 0.265 & 0.265 & 0.265 & 0.265 \\
EAT-D & 0.255 & 0.255 & 0.249 & 0.243 & 0.245 & 0.255 & 0.254 & 0.253 & 0.251 & 0.250 & 0.261 & 0.265 & 0.264 & 0.262 & 0.262 \\
EAT-DA & 0.260 & 0.261 & 0.263 & 0.264 & 0.263 & 0.270 & 0.269 & 0.270 & 0.269 & 0.270 & 0.269 & 0.269 & 0.270 & 0.269 & 0.270 \\
PPO & 0.228 & 0.228 & 0.228 & 0.228 & 0.228 & 0.228 & 0.228 & 0.228 & 0.228 & 0.228 & 0.228 & 0.228 & 0.228 & 0.228 & 0.228 \\
Genetic & 0.202 & 0.216 & 0.188 & 0.189 & 0.203 & 0.201 & 0.224 & 0.224 & 0.219 & 0.219 & 0.199 & 0.224 & 0.222 & 0.223 & 0.223 \\
Harmony & 0.214 & 0.214 & 0.204 & 0.198 & 0.203 & 0.185 & 0.196 & 0.196 & 0.202 & 0.202 & 0.211 & 0.183 & 0.177 & 0.177 & 0.211 \\
Random & 0.197 & 0.193 & 0.192 & 0.194 & 0.195 & 0.193 & 0.192 & 0.190 & 0.200 & 0.187 & 0.189 & 0.190 & 0.197 & 0.190 & 0.186 \\
Greedy & 0.270 & 0.270 & 0.270 & 0.270 & 0.270 & 0.270 & 0.270 & 0.270 & 0.270 & 0.270 & 0.270 & 0.270 & 0.270 & 0.270 & 0.270 \\

\end{tabular}
}
\end{table*}

\begin{table*}
\centering
\caption{Response Latency}
\label{tb:response_time}
\resizebox{\textwidth}{!}{
\begin{tabular}{l|rrrrr|rrrrr|rrrrr}
\multirow{3}{*}{Algorithm} & \multicolumn{15}{c}{Task Arrival Rate}                                                                         \\ 
\cline{2-16}
                           & \multicolumn{5}{c|}{4 Nodes}        & \multicolumn{5}{c|}{8 Nodes}         & \multicolumn{5}{c}{12 Nodes}      \\ 
\cline{2-16}
                           & 0.01 & 0.03 & 0.05  & 0.07  & 0.09  & 0.06 & 0.08  & 0.1   & 0.12  & 0.14  & 0.11 & 0.13 & 0.15 & 0.17 & 0.19  \\ 
\hline
{EAT} & {30.3} & {21.4} & {39.7} & {52.5} & {100.9} & {50.0} & {56.9} & {76.1} & {84.0} & {91.1} & {27.2} & {36.5} & {38.7} & {36.7} & {41.5} \\
EAT-A                      & 32.8 & 21.0 & 55.7  & 79.8  & 101.1 & 49.6 & 73.3  & 89.9  & 100.0 & 102.9 & 36.0 & 41.0 & 41.0 & 55.4 & 43.0  \\
EAT-D                       & 72.8 & 64.2 & 111.1 & 135.5 & 158.4 & 70.9 & 69.0  & 101.0 & 122.3 & 121.7 & 42.8 & 43.0 & 43.1 & 49.9 & 48.6  \\
EAT-DA                        & 57.0 & 39.1 & 95.0  & 133.0 & 170.3 & 74.3 & 79.7  & 96.0  & 110.0 & 131.2 & 46.0 & 46.3 & 46.1 & 50.3 & 47.2  \\
PPO                        & 65.0 & 71.7 & 127.3 & 147.1 & 155.9 & 80.0 & 84.2  & 105.0 & 96.8  & 96.8  & 45.4 & 44.3 & 42.7 & 49.4 & 49.4  \\
Genetic                         & 36.5 & 36.4 & 43.4  & 46.5  & 70.5  & 35.5 & 37.0  & 37.0  & 40.5  & 40.5  & 25.6 & 24.7 & 24.6 & 26.6 & 25.9  \\
Harmony                         & 41.3 & 39.4 & 51.6  & 58.4  & 78.7  & 41.1 & 33.7  & 33.7  & 39.2  & 39.2  & 29.9 & 24.8 & 23.2 & 26.1 & 29.3  \\
Random                        & 47.9 & 48.7 & 56.1  & 57.9  & 62.4  & 49.0 & 51.5  & 52.5  & 53.4  & 59.3  & 30.2 & 35.1 & 34.7 & 33.6 & 37.9  \\
Greedy                     & 64.1 & 86.6 & 154.2 & 176.1 & 182.8 & 91.7 & 105.6 & 117.2 & 128.6 & 131.1 & 51.4 & 53.5 & 53.9 & 59.9 & 56.3 \\

\end{tabular}
}
\end{table*}

\begin{table*}
\centering
\caption{Reload Rate}
\label{tb:reload_rate}
\resizebox{\textwidth}{!}{
\begin{tabular}{l|rrrrr|rrrrr|rrrrr}
\multirow{3}{*}{Algorithm} & \multicolumn{15}{c}{Task Arrival Rate}                                                                                                \\ 
\cline{2-16}
                           & \multicolumn{5}{c|}{4 Nodes}               & \multicolumn{5}{c|}{8 Nodes}               & \multicolumn{5}{c}{12 Nodes}                \\ 
\cline{2-16}
                           & 0.01 & 0.03 & 0.05 & 0.07 & 0.09 & 0.06 & 0.08 & 0.1 & 0.12 & 0.14 & 0.11 & 0.13 & 0.15 & 0.17 & 0.19  \\ 
\hline
{EAT} & {0.605} & {0.547} & {0.633} & {0.594} & {0.625} & {0.761} & {0.634} & {0.604} & {0.570} & {0.651} & {0.375} & {0.430} & {0.328} & {0.411} & {0.430} \\
EAT-A & 0.629 & 0.573 & 0.667 & 0.636 & 0.605 & 0.719 & 0.653 & 0.695 & 0.617 & 0.581 & 0.473 & 0.430 & 0.386 & 0.419 & 0.460 \\
EAT-D & 0.673 & 0.717 & 0.673 & 0.791 & 0.639 & 0.844 & 0.756 & 0.706 & 0.694 & 0.713 & 0.556 & 0.537 & 0.506 & 0.531 & 0.534 \\
EAT-DA & 0.734 & 0.675 & 0.700 & 0.700 & 0.794 & 0.774 & 0.750 & 0.719 & 0.719 & 0.833 & 0.533 & 0.519 & 0.566 & 0.512 & 0.591 \\
PPO & 0.723 & 0.675 & 0.655 & 0.736 & 0.697 & 0.812 & 0.750 & 0.688 & 0.734 & 0.713 & 0.625 & 0.531 & 0.487 & 0.531 & 0.562 \\
Genetic & 0.761 & 0.813 & 0.850 & 0.654 & 0.767 & 0.769 & 0.846 & 0.846 & 0.897 & 0.917 & 0.635 & 0.521 & 0.521 & 0.593 & 0.614 \\
Harmony & 0.739 & 0.819 & 0.726 & 0.717 & 0.873 & 0.917 & 0.846 & 0.806 & 0.844 & 0.814 & 0.700 & 0.557 & 0.545 & 0.614 & 0.700 \\
Random & 0.775 & 0.868 & 0.802 & 0.786 & 0.905 & 0.887 & 0.909 & 0.870 & 0.843 & 0.854 & 0.648 & 0.626 & 0.586 & 0.564 & 0.602 \\
Greedy & 0.551 & 0.542 & 0.586 & 0.504 & 0.516 & 0.594 & 0.578 & 0.578 & 0.527 & 0.550 & 0.468 & 0.411 & 0.398 & 0.403 & 0.448 \\
\end{tabular}
}
\end{table*}

\subsubsection{Baselines}

We compare EAT against these baselines:

\begin{enumerate}
    \item \textbf{EAT-DA} \cite{b4.2}: The baseline method, which is the standard SAC algorithm. This corresponds to the full EAT algorithm with both the attention and diffusion components removed.
    \item \textbf{EAT-D}: An ablation of the EAT algorithm that removes the diffusion component. It enhances the SAC method with an attention mechanism.
    \item \textbf{EAT-A} \cite{b2.3}: An ablation of the EAT algorithm that removes the attention component. It enhances the SAC method by incorporating a diffusion model.
    \item \textbf{PPO} \cite{schulman2017ppo}: Proximal Policy Optimization (PPO) serves as an on-policy DRL method.
    \item \textbf{Harmony} \cite{geem2001new}: Harmony Search (Harmony) is a computationally efficient metaheuristic optimization algorithm.
    \item \textbf{Genetic} \cite{holland1992adaptation}: Genetic Algorithm (Genetic) is a computationally intensive metaheuristic optimization method.
    \item \textbf{Random}: It Randomly selects an action and adopts the Task selector and Server selector (Random) to allocate the task.
    \item \textbf{Greedy}: It selects actions to maximize immediate rewards by evaluating all policies.
\end{enumerate}

\subsection{Experimental Results}

\subsubsection{Image Generation Examples}

The demonstration of the generation results are shown in Fig.~\ref{demo}. We select a subset of image results produced by our implemented system with the scheduling of the EAT algorithm. In the experiments, we select image generation tasks for five distinct prompts, submitting jobs that partitioned each image into 1, 2, or 4 patches, and recorded the average execution times and quality scores. As shown in the figure, the 4-patch configuration exhibits slight visual differences due to more partitioning, yet without any appreciable degradation in overall image quality. In terms of execution time, the average latency decreased markedly with increased parallelism: compared to the original group, the 2-patch and 4-patch tasks achieve speedups of 1.63$\times$ and 2.07$\times$, respectively.

\subsubsection{Training Results}

\begin{figure*}[!t]
  \centering
  \begin{subfigure}[b]{0.31\linewidth}
    \centering
    \includegraphics[width=\linewidth]{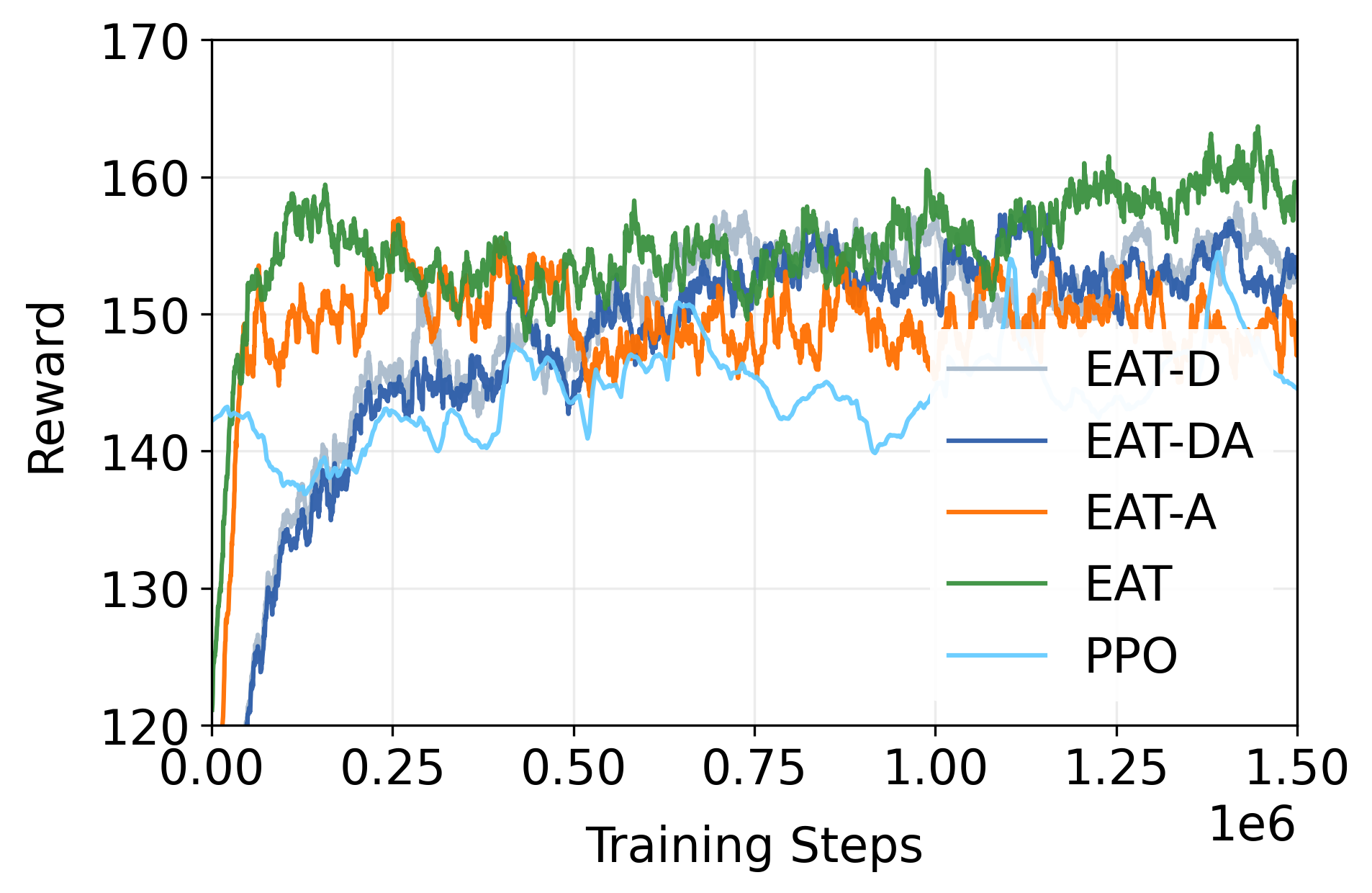}
    \caption{Reward}
    \label{fig:training_reward}
  \end{subfigure}
  \begin{subfigure}[b]{0.31\linewidth}
    \centering
    \includegraphics[width=\linewidth]{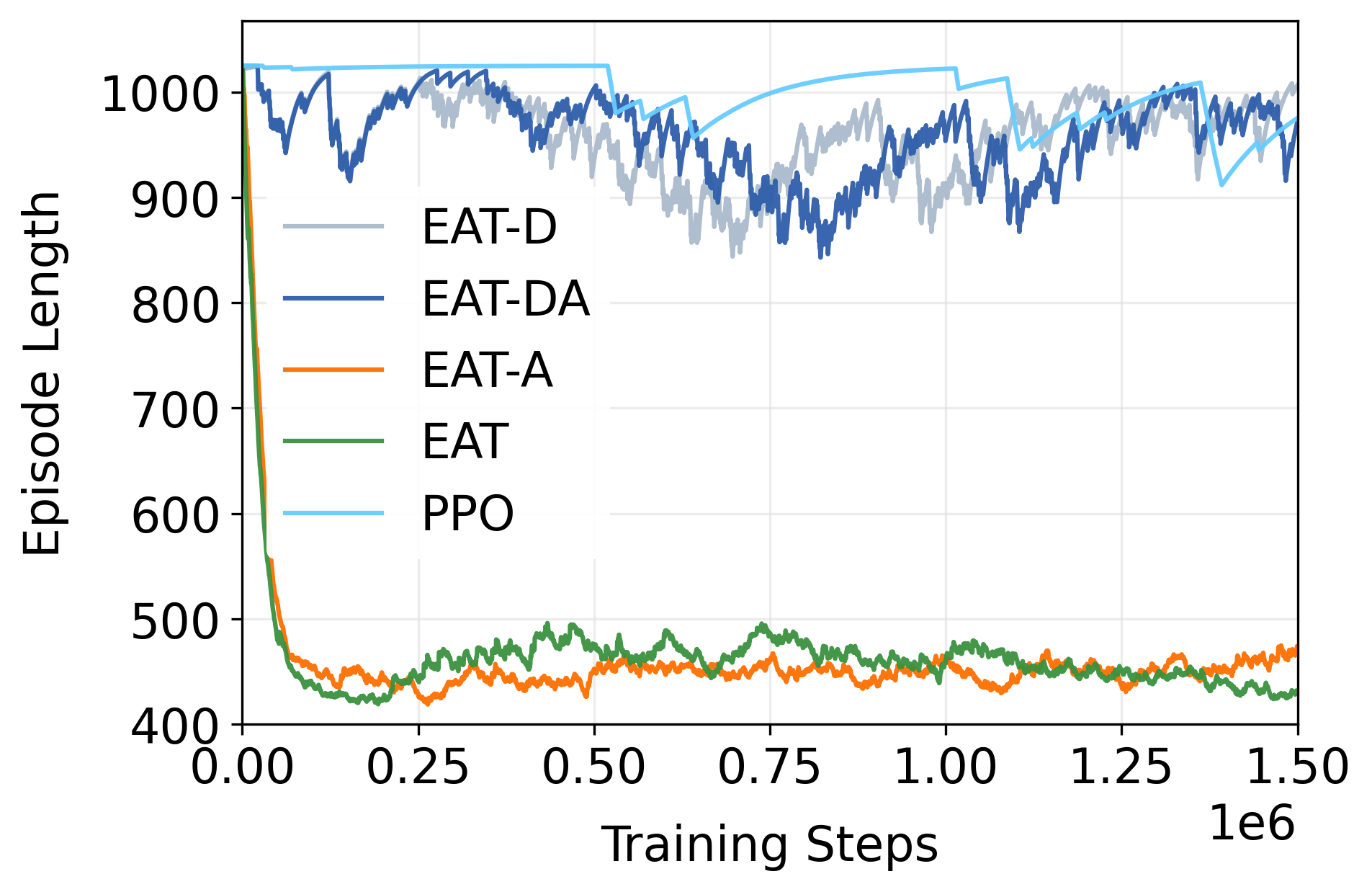}
    \caption{Episode Length}
    \label{fig:training_length}
  \end{subfigure}
  \hfill
  \begin{subfigure}[b]{0.31\linewidth}
    \centering
    \includegraphics[width=\linewidth]{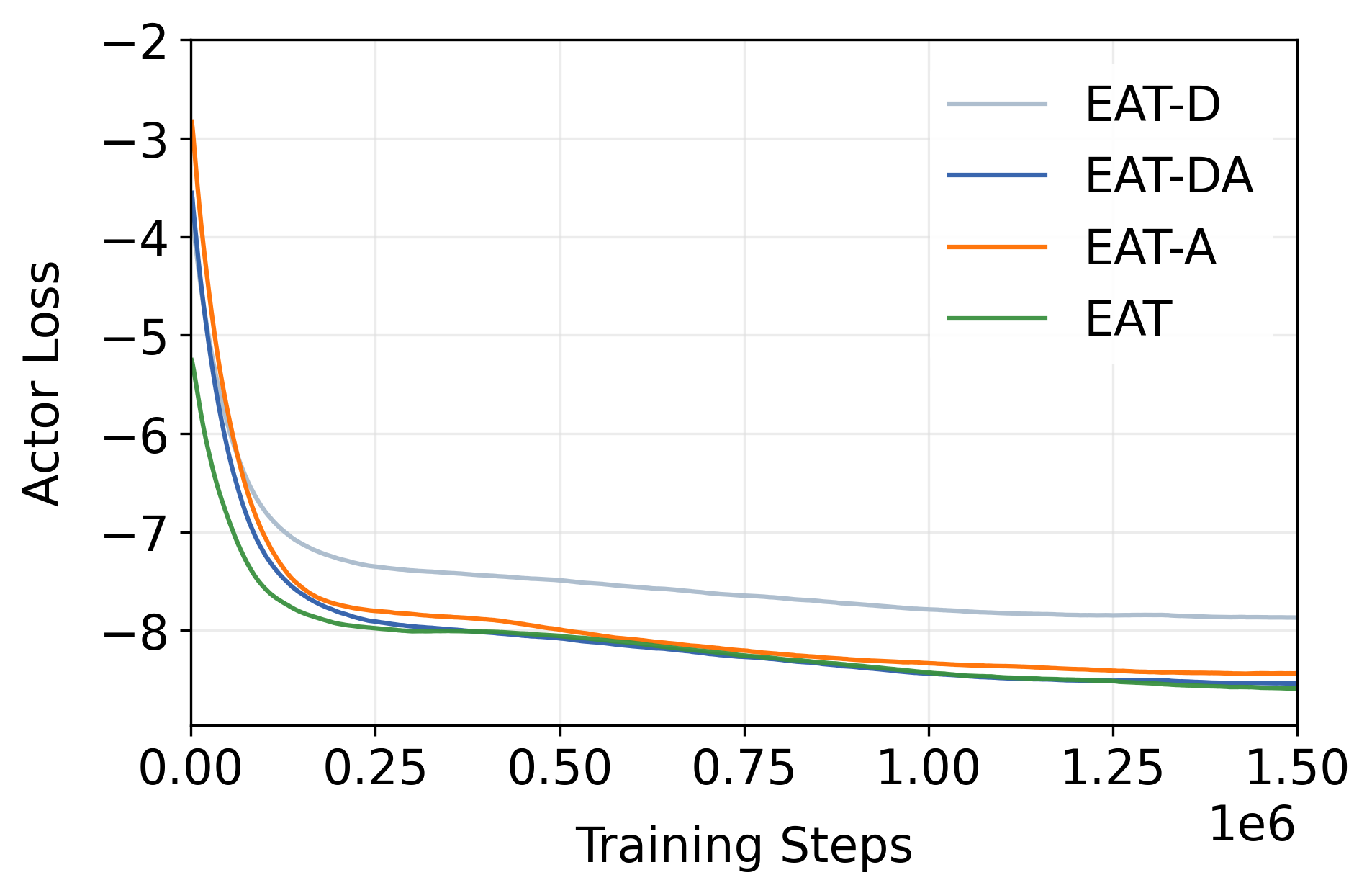}
    \caption{Loss}
    \label{fig:training_loss}
  \end{subfigure}
  \hfill
  \caption{Training Metrics in 8 Servers Environment.}
  \label{fig:training}
\end{figure*}

We evaluate training rewards, policy network loss, and episode lengths of DRL algorithms in the 8-server environment, as shown in Fig.~\ref{fig:training}. After $7.5\times 10^5$ steps, rewards of EAT-DA, EAT and EAT-A curve stabilized, with the fluctuation of PPO. Fig.~\ref{fig:training_length} shows that EAT-A and EAT converge rapidly to approximately $450$ steps. EAT-DA and PPO often fail to stay within the step limit, indicating poor adaptability and resulting in excessively long episodes. EAT-DA struggles to generate effective actions, hindering task completion and episode termination. PPO, prone to local optima due to its policy gradient approach, exhibits even poorer episode length performance than EAT-DA in the testing environment. EAT maintains an upward trend and outperforms others, while EAT-A and EAT-DA show no significant improvement as shown in Fig.~\ref{fig:training_reward}.

\subsubsection{Quality Performance}

The TABLE.~\ref{tb:quality} compares task quality across algorithms in a real-world 4-server setup and simulated 8-server and 12-server environments. The Greedy algorithm maximizes inference steps for slight quality advantage, but it increases latency dramatically. In contrast, DRL-based algorithms show minimal performance differences and slightly lower quality, as they prioritize faster inference latency at the expense of some quality to avoid delaying subsequent tasks. However, the PPO easily gets stuck in local optima when making step decisions. In different environments, PPO tends to find a fixed step to balance time and quality, causing low and fixed quality performance. Conversely, the meta-heuristic methods precompute a fixed action sequence to maximize the reward, which leads to poor performance in highly dynamic environments, resulting in lower quality. The Random algorithm leads to significantly lower quality due to its reduced average inference step value. This reduction in inference steps notably decreases quality, with task performance ranked as Greedy $>$ SAC-based methods $>$ PPO $>$ meta-heuristic methods $>$ Random.

\subsubsection{Inference Latency Performance}


The Table.~\ref{tb:response_time} compares the inference latencies of various algorithms across different environments. The EAT algorithm excels by reusing model initialization results and dynamically adjusting task inference latencies. In contrast, the Greedy approach prioritizes task quality, leading to higher per-task latency that accumulates, significantly increasing overall inference latencies. The EAT-DA algorithm struggles to prioritize tasks with longer queue times, resulting in prolonged inference latencies and occasional task stalling, as shown in Fig.~\ref{fig:training_length}. PPO, despite its susceptibility to local optima, achieves response times comparable to EAT-DA due to its low-step policy. Meta-heuristic methods and Random similarly benefit from low-step policies, resulting in relatively low response times. In a real-world 4-server experiment, the EAT algorithm outperforms the EAT-A by 16 (28.7\%), EAT-DA by 55.3 (58.2\%),  PPO by 127.3 (68.8\%), Greedy by 114.5 (74.3\%), Random by 16.4 (30.0\%), Genetic by 3.6 (8.3\%), and Harmony by 51.6 (23.1\%).



\subsubsection{Model Reload Rate Analysis}


The TABLE \ref{tb:reload_rate} shows model reload rates across algorithms in 4-server real-world and 8/12-server simulated environments. Lower reload rates reflect better server resource efficiency by reducing redundant model reloads. EAT achieves the overall lowest reload rate, while EAT-DA and PPO perform similarly but less efficiently. PPO often converges to local optima, performing well at a 0.1 task arrival rate but poorly at others. Due to the longer execution time, the Greedy algorithm causes significant task backlog, resulting in a larger pool of available tasks for selection, which helps avoid reinitialization. In smaller-scale scenarios, it exhibits a lower reload rate. However, in larger-scale environments, tasks are completed more quickly, and the reload rate gradually approaches the EAT. Random has overall the highest reload rate due to its random task selection and disregard for pre-loaded models. Metaheuristic algorithms, lacking environmental feedback, often perform similarly to random algorithms. In the real-world 4-server experiment, EAT achieves a reload rate of $0.633$, outperforming EAT-A ($0.667$), EAT-DA ($0.700$), PPO ($0.688$), Harmony($0.726$), Genetic($0.850$), and Random ($0.800$).

\subsubsection{Initialization Overhead Analysis}

\begin{figure*}[t] \centering \includegraphics[width=0.96\linewidth]{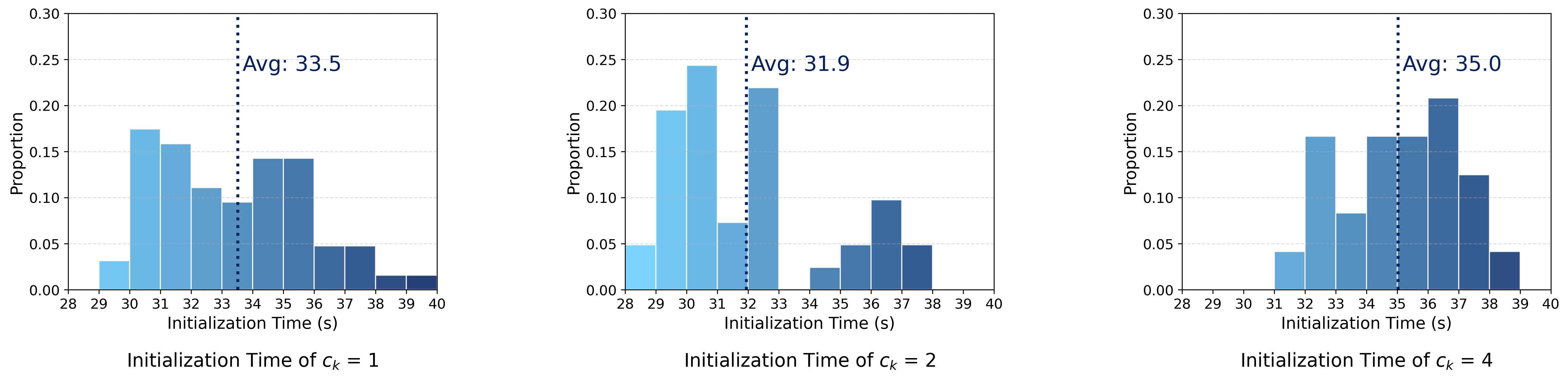} \caption{Initialization Time with Different Cooperate Number.} \label{loading_time} \end{figure*}

Fig.~\ref{loading_time} highlights unpredictable variations in initialization times. Real-world testing showed EAT's remarkable resilience to temporal fluctuations, surpassing all baselines. This advantage arises from EAT's attention mechanism, which adeptly captures the server information, outperforming the MLP architectures in EAT-A and EAT-DA that struggle with information interference.

\begin{figure*}[t] \centering \includegraphics[width=0.96\linewidth]{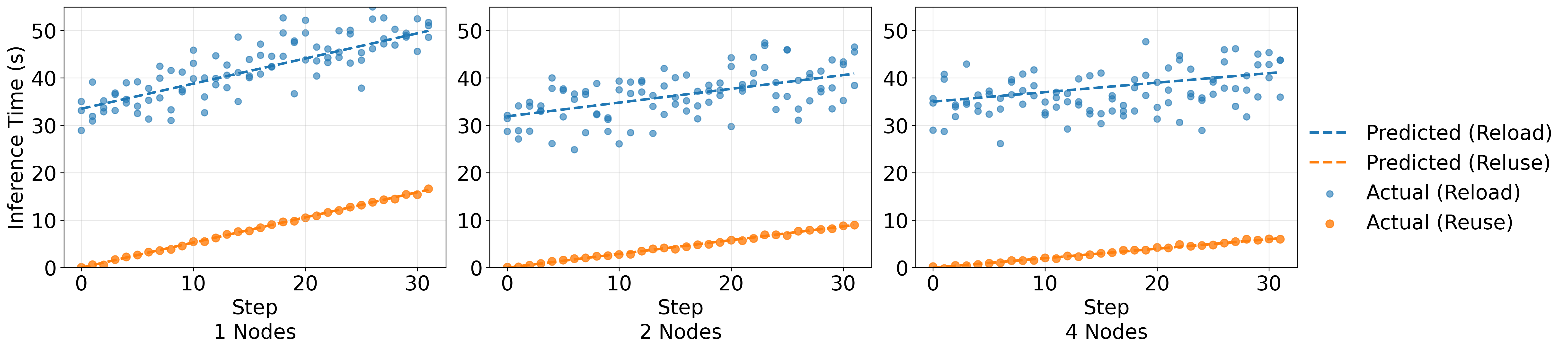} \caption{Time Prediction with Different Cooperate Number.} \label{fig:time_prediciton} \end{figure*}

\begin{figure*}[t] \centering \includegraphics[width=0.96\linewidth]{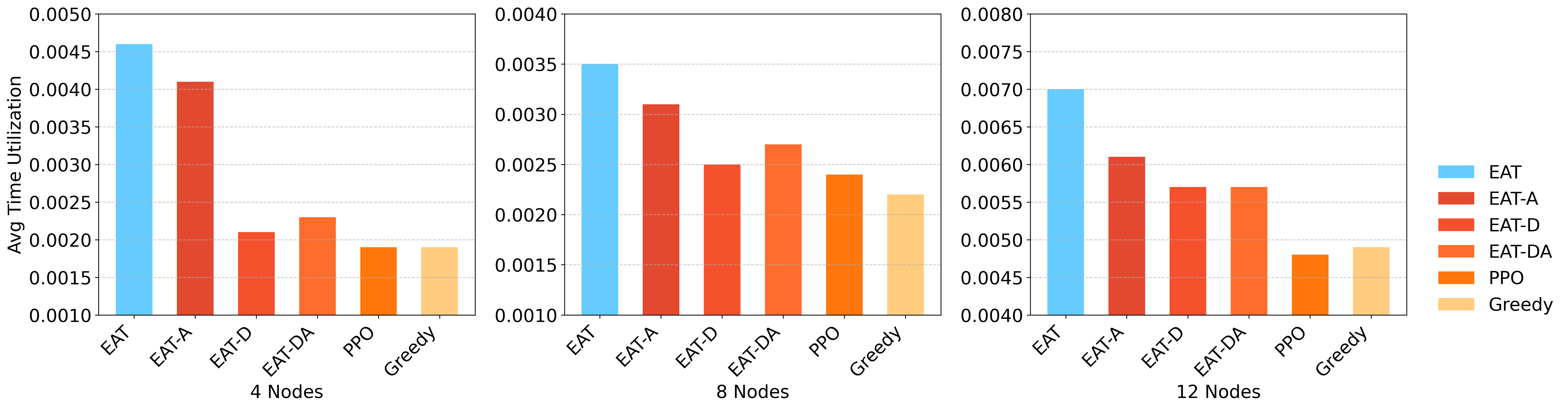} \caption{Generation Efficiency.} \label{fig:efficiency} \end{figure*}

\subsubsection{Time Prediction Analysis}
Fig.~\ref{fig:time_prediciton} shows the time prediction for tasks with and without model reloading. For tasks that do not require model reloading, execution time grows linearly with the number of draw steps, as stable diffusion's computational requirements are proportional to step count. As the number of collaborative nodes increases, computation speed progressively improves. For tasks requiring model reloading, execution times show randomness due to unstable loading times, but generally fall around the time predictor's estimates. While the time predictor is less accurate for these reloading tasks, it still adequately reflects node load conditions, which is sufficient for scheduling purposes.

\subsubsection{Efficiency Analysis}
Fig.~\ref{fig:efficiency} demonstrates the comprehensive performance of different algorithms. We calculate efficiency as the ratio of quality to response latency, measuring the quality achieved per second to evaluate overall efficiency. Specifically, algorithms employing diffusion models exhibit superior performance compared to other RL methods, with the EAT algorithm particularly excelling due to its attention mechanism, which maintains higher time utilization compared to EAT-A even in large-scale experiments, while EAT-D, EAT-DA, and PPO methods show no significant performance differences among themselves. The Greedy method demonstrates lower efficiency due to excessive draw steps, investing substantial time without proportional quality improvements. Random and meta-heuristic algorithms are excluded from this analysis as their image quality falls below the basic quality requirement required for text-to-image generation tasks. The time utilizaiton ranked as: EAT $>$ EAT-A $>$ EAT-DA $>$ EAT-D $>$ PPO $>$ Greedy.

\subsubsection{Inference Latency}
As shown in TABLE \ref{tab:algo_time}, the Greedy algorithm incurs the highest inference latency due to exhaustive enumeration of all possible actions. As the maximum number of drawing steps and the number of servers increase, its latency grows correspondingly. EAT and EAT-A exhibit relatively higher latency, primarily due to repeated inference through the denoise net. Additionally, EAT and EAT-D introduce extra computation time compared to EAT-A and EAT-DA, respectively, as a result of incorporating the attention mechanism. Meta-heuristic methods require no inference time since they rely on pre-generated action sequences. The Random algorithm also exhibits near-zero latency. While inference times vary across algorithms, all methods perform acceptably within the millisecond range.

\subsubsection{Ablation Analysis}
The EAT-A algorithm lacks the attention layer present in EAT, while EAT-D omits the diffusion component, and EAT-DA is based on the foundational EAT-DA framework. In terms of task quality, all four algorithms perform similarly. Regarding reload rates, EAT-A exhibits higher reload rates due to its absence of the attention layer, which reduces its feature extraction capability, leading to higher response times. EAT-D, lacking the diffusion layer, struggles to generate high-quality actions, resulting in performance that is close to EAT-DA. Notably, EAT-D's improvement over EAT-DA is smaller than EAT's improvement over EAT-A, highlighting the significant role of the diffusion process. By gradually denoising, diffusion effectively amplifies the advantages brought by high-quality features.

\begin{table}[t]
    \begin{center}
    \caption{Inference Latency}\label{tab:algo_time}
    \begin{tabular}{l|r}
      \hline
      \textbf{Algorithm} & \textbf{Time (s)} \\
      \hline
      Greedy & $2.38 \times 10^{-2}$ \\
      EAT  & $1.12 \times 10^{-2}$ \\
      EAT-A  & $9.88 \times 10^{-3}$ \\
      EAT-DA   & $1.18 \times 10^{-3}$ \\
      PPO   & $1.08 \times 10^{-3}$ \\
      Random & $\approx 0$ \\
      Genetic    & $\approx 0$ \\
      Harmony    & $\approx 0$ \\
      \hline
    \end{tabular}
    \end{center}
\end{table}

\section{Discussion}

In our experiments, we have measured the communication latency for transferring a generated 3MB image. The average latency between two physical servers is 0.175 seconds, and between a container and its host on a single server is 0.031 seconds. These values are negligible compared to the overall task response time of several seconds, due to our system’s asynchronous design. Image transmission to the scheduler and the next image generation occur concurrently, effectively hiding network delays and preventing them from becoming a bottleneck. Therefore, network latency has minimal impact on overall system performance and response time in our settings.

Moreover, a key finding from our experiments is that image generation involves significant response times, primarily due to the unoptimized model loading process in DistriFusion. To alleviate this bottleneck, future work could focus on more efficient model caching. For sequential tasks using the same model parameters, rebuilding only the communication process group for new servers—rather than fully unloading and reloading the model—would reduce initialization overhead and improve overall response times.

\section{Conclusion} \label{Conclusion}

In this paper, we presented EAT, a QoS-aware edge-collaborative scheduling algorithm designed to balance inference latency and quality for multi-distributed AIGC models. By leveraging gang scheduling concepts, an attention layer for resource-aware feature extraction, and a diffusion-based policy network, our EAT algorithm flexibly segments and schedules AIGC tasks across diverse edge servers. We have open-sourced the entire EAT framework code and validated its effectiveness through real systems and large-scale simulations. Our results show that EAT reduces inference latency by up to 74.3\% while maintaining high QoS. Future work will expand this approach by investigating horizontal scaling for multi-modal AIGC tasks, integrating temporal adaptation for dynamic edge environments, and optimizing energy consumption in distributed inference.

{\footnotesize
\bibliographystyle{IEEEtran}
\bibliography{references}
}

\begin{IEEEbiography}[{\includegraphics[width=1in,height=1.25in,clip,keepaspectratio]{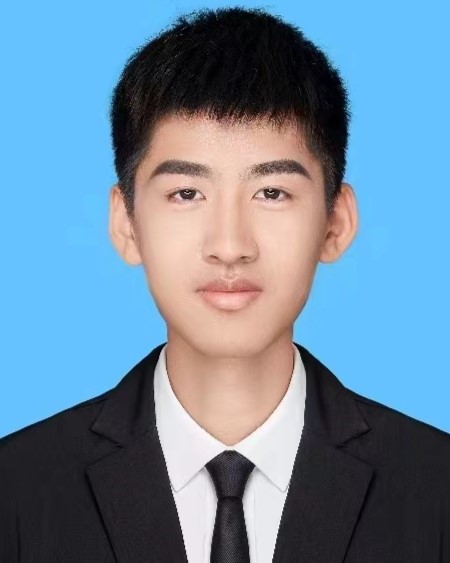}}]{Zhifei Xu}
is currently pursuing a bachelor’s degree in the Faculty of Arts and Sciences, Beijing Normal University at Zhuhai, China. His main research interests include edge computing, reinforcement learning and their applications.
\end{IEEEbiography}

\begin{IEEEbiography}[{\includegraphics[width=1in,height=1.25in,clip,keepaspectratio]{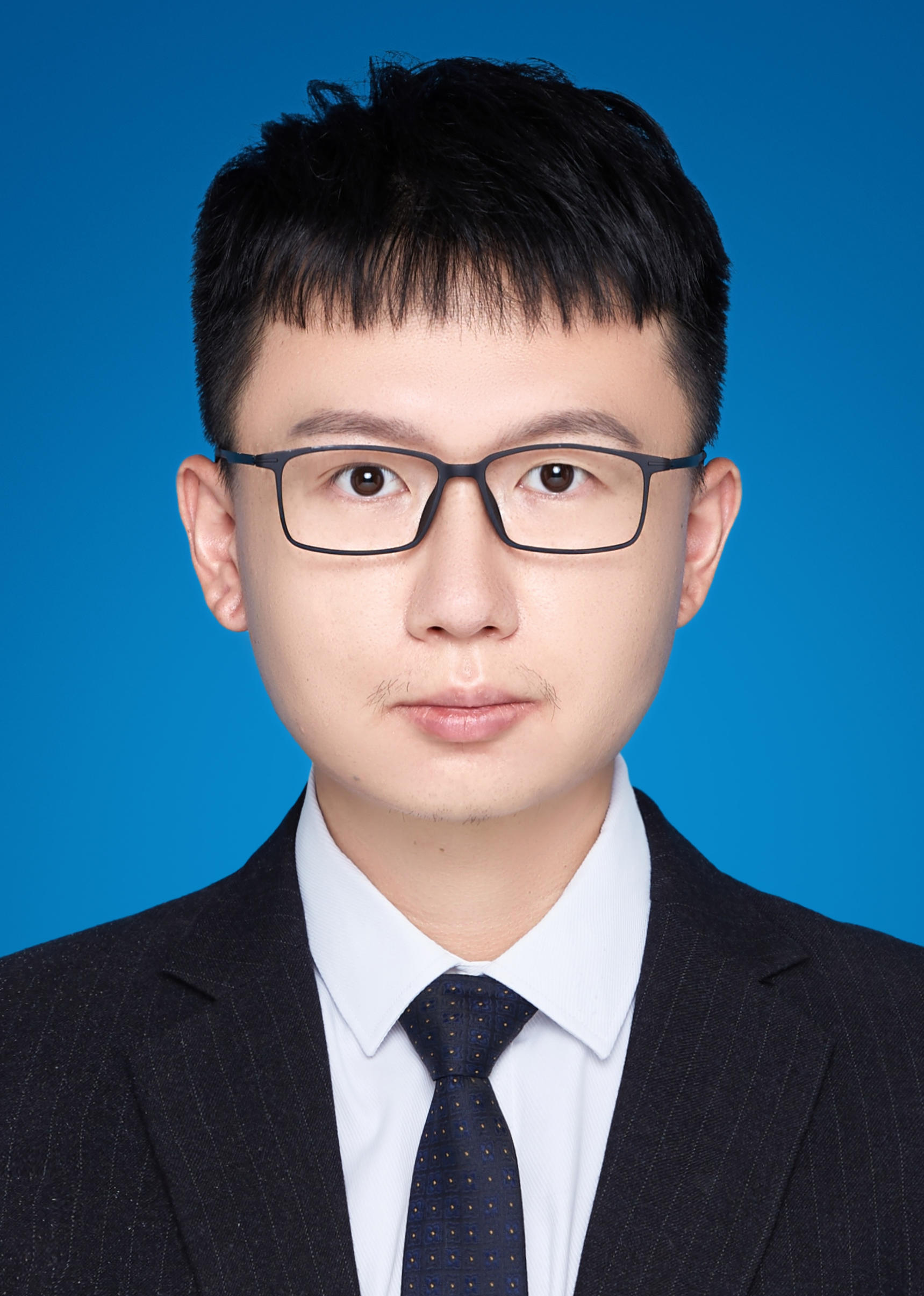}}]{Zhiqing Tang}
        received the B.S. degree from School of Communication and Information Engineering, University of Electronic Science and Technology of China, China, in 2015 and the Ph.D. degree from Department of Computer Science and Engineering, Shanghai Jiao Tong University, China, in 2022. He is currently an Assistant Professor with the Advanced Institute of Natural Sciences, Beijing Normal University, China. His current research interests include edge computing, resource scheduling, and reinforcement learning.
\end{IEEEbiography}

\begin{IEEEbiography}[{\includegraphics[width=1in,height=1.25in,clip,keepaspectratio]{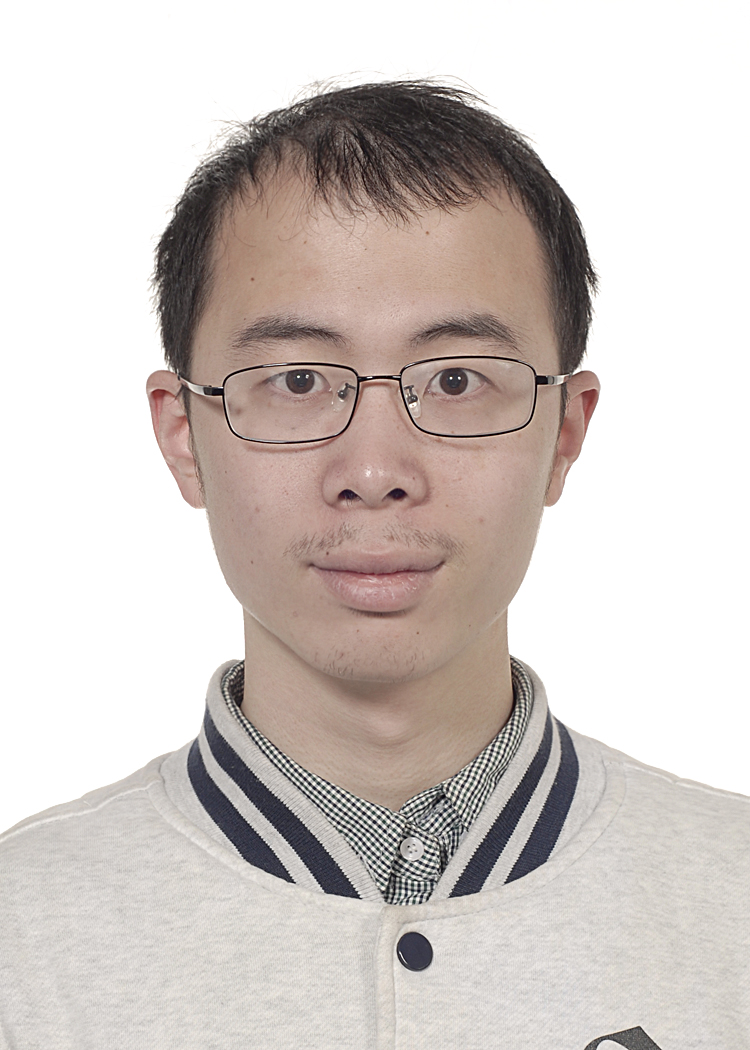}}]{Jiong Lou}
	received the B.S. degree and Ph.D. degree in the Department of Computer Science and Engineering, Shanghai Jiao Tong University, China, in 2016 and 2023. Since 2023, he has held the position of Research Assistant Professor in the Department of Computer Science and Engineering, Shanghai Jiao Tong University, China. He has published more than ten papers in leading journals and conferences (e.g., ToN, TMC and TSC). His current research interests include edge computing, task scheduling and container management.
\end{IEEEbiography}

\begin{IEEEbiography}[{\includegraphics[width=1in,height=1.25in,clip,keepaspectratio]{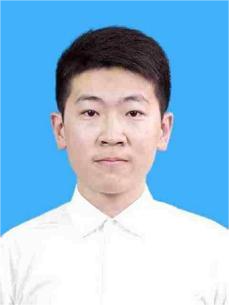}}]{Zhi Yao} received the B.S. degree from College of Electronic and Information Engineering, Shandong University of Science and Technology, China, in 2020, and the M.S. degree from South China Academy of Advanced Optoelectronics, South China Normal University, China, in 2023. He is currently pursuing the Ph.D. degree in School of Artificial Intelligence, Beijing Normal University, China. His current research interests include mobile edge computing, vector database, LLM request scheduling, and reinforcement learning.
\end{IEEEbiography}

\begin{IEEEbiography}[{\includegraphics[width=1in,height=1.25in,clip,keepaspectratio]{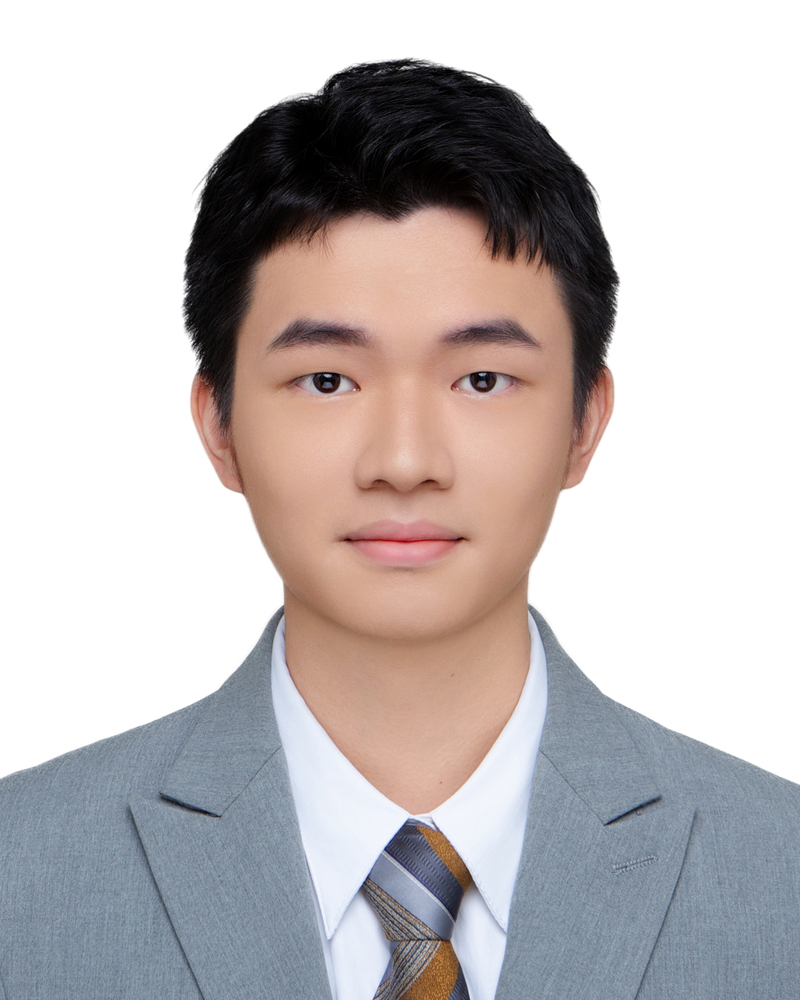}}]{Xuan Xie}
is currently pursuing a bachelor’s degree in the Faculty of Arts and Sciences, Beijing Normal University at Zhuhai, China. His main research interests include mobile edge computing, edge intelligence, and their applications.
\end{IEEEbiography}

\begin{IEEEbiography}
[{\includegraphics[width=1in,height=1.25in,clip,keepaspectratio]{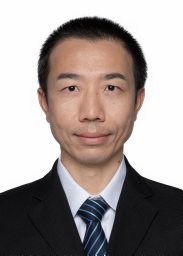}}]{Tian Wang}(Senior Member, IEEE) received his BSc and MSc degrees in Computer Science from the Central South University in 2004 and 2007, respectively. He received his PhD degree in City University of Hong Kong in Computer Science in 2011. Currently, he is a Professor in the Institute of AI and Future Networks, Beijing Normal University. His research interests include internet of things, edge computing, and mobile computing. He has 27 patents and has published more than 200 papers in high-level journals and conferences. He has more than 16000 citations and his H-index is 73. 
\end{IEEEbiography}

\begin{IEEEbiography}
[{\includegraphics[width=1in,height=1.25in,clip,keepaspectratio]{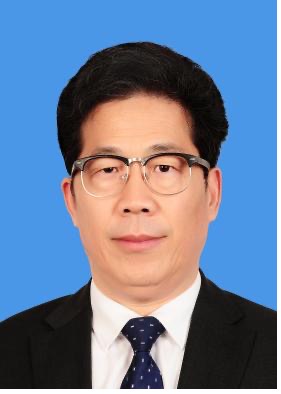}}]{Yinglong Wang}(Senior Member, IEEE) is currently the director of the Key Laboratory of Computing Power Network and Information Security, Ministry of Education. He received his PhD degree in Shandong University, China, in 2005. His current research interests include edge computing,  network communication and data science. He has undertaken more than 20 national projects, won 3 first prizes of provincial science and technology awards, compiled 10 national standards, published 3 monographs, and published more than 120 high-level papers. 
\end{IEEEbiography}

\begin{IEEEbiography}[{\includegraphics[width=1in,height=1.25in,clip,keepaspectratio]{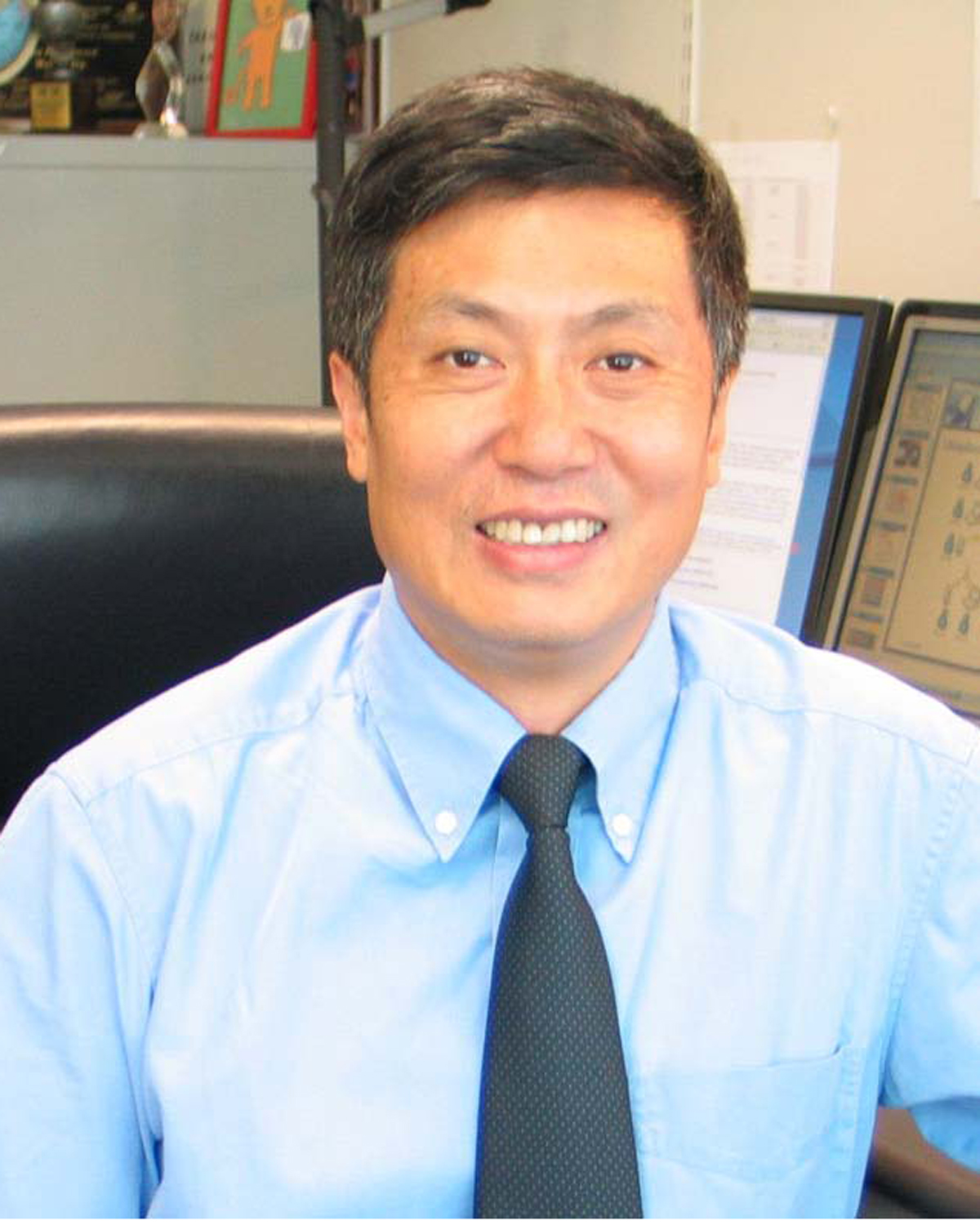}}]{Weijia Jia}(Fellow, IEEE) is currently the Director of Institute of AI and Future Networks, and the Director of Super Intelligent Computer Center, Beijing Normal University; also a Chair Professor at UIC, China. He has served as the VP for Research at UIC in 2020-2024. Prior joining BNU, he served as the Deputy Director of State Key Lab of IoT for Smart City at University of Macau and Zhiyuan Chair Professor at Shanghai Jiaotong University. From 95-13, he worked in City University of Hong Kong as a professor. He has over 700 publications in international journals/conferences and research books. He has received the 1st Prize of Scientific Research Awards from the Ministry of Education of China in 2017 (list 2), and top 2\% World Scientists in Stanford-list (2020-2024). He is the Fellow of IEEE and the Distinguished Member of CCF.
\end{IEEEbiography}

\end{document}